\documentclass{aa}  

\usepackage{graphicx}
\usepackage{txfonts}
\usepackage{amsmath}
\usepackage[normalem]{ulem}
\usepackage{hyperref}

\def\rsun{{~R}_{\odot}}
\def\msun{{~M}_{\odot}}
\def\zsun{{~Z}_{\odot}}
\def\logp{{\rm log} \, P}
\def\logpday{{\rm log} \, P \, {\rm(days)}}
\def\gpy{{\rm ~Gpc}^{-3} {\rm ~yr}^{-1}}

\newcommand{\kms}{\ensuremath{\,\rm{km}\,\rm{s}^{-1}}}

\newcommand{\Msun}{\ensuremath{\,M_\odot}}

\usepackage{color}
\bibpunct[; ]{(}{)}{;}{a}{}{;}

\defcitealias{md16}{MD17}
\defcitealias{dMB15}{dMB15}
\defcitealias{B16Nat}{B16}

\begin{document} 

   \title{Impact of inter-correlated initial binary parameters on double black hole and neutron star mergers} 

   \author{J. Klencki\inst{1}
          \and
          M. Moe\inst{2,3} 
          \and
          W. Gladysz\inst{4}
          \and
          M. Chruslinska\inst{1}
          \and
          D. E. Holz\inst{5,6}
          \and
          K. Belczynski\inst{7}               
   }

   \institute{
   Department of Astrophysics/IMAPP, Radboud University, P O Box 9010, NL-6500 GL Nijmegen, The Netherlands\\
          \email{j.klencki@astro.ru.nl}
         \and
           Steward Observatory, University of Arizona, 
           933 N. Cherry Ave., Tucson, AZ 85721, USA
         \and
           Einstein Fellow
         \and 
         Astronomical Observatory, Warsaw University, Ujazdowskie 4, 00-478 Warsaw, Poland
         \and
           Enrico Fermi Institute, Department of Physics, Department of
           Astronomy and Astrophysics, and Kavli Institute for Cosmological Physics, 
           University of Chicago, Chicago, IL 60637, USA
         \and
         Kavli Institute for Particle Astrophysics \& Cosmology and Physics Department,
         Stanford University , Stanford, CA 94305, USA
         \and
         Nicolaus Copernicus Astronomical Center, Polish Academy of Sciences,
           ul. Bartycka 18, 00-716 Warsaw, Poland        
   }

   \date{Received Sep 12, 2016; accepted ???}

  \abstract
   {
    The distributions of the initial main-sequence binary parameters are one of the key
    ingredients in obtaining evolutionary predictions for compact
    binary (BH-BH\,/\,BH-NS\,/\,NS-NS) merger rates.
    Until now, such calculations were done under the assumption that initial
    binary parameter distributions were independent. For the first time,
    we implement empirically derived inter-correlated
    distributions of initial binary parameters primary mass ($M_1$), mass ratio ($q$),
    orbital period ($P$), and eccentricity ($e$).

    Unexpectedly, the introduction of inter-correlated initial binary
    parameters leads to only a small decrease in the predicted merger
    rates by a factor of $\lesssim 2-3$ relative
    to the previously used non-correlated initial distributions.
    The formation of compact object mergers in the isolated classical binary evolution
    favours initial binaries with stars of comparable masses ($q\approx$\,0.5\,$-$\,1) at
    intermediate orbital periods ($\logpday$ = 2\,$-$\,4). New distributions slightly
    shift the mass ratios towards lower values with respect to the previously used flat
    $q$ distribution, which is the dominant effect decreasing the rates. 
    New orbital periods ($\sim$ 1.3 more initial systems within $\logpday$ = 2\,$-$\,4), together
    with new eccentricities (higher), only negligibly increase the number of progenitors
    of compact binary mergers.

    Additionally, we discuss the uncertainty of merger rate predictions associated 
    with possible variations of the massive-star initial mass function (IMF). 
    We argue that evolutionary calculations should be normalized to a star
    formation rate (SFR) that is obtained from the observed amount of UV light
    at wavelength 1500$\AA$ (an SFR indicator). In this case, contrary to recent reports,
    the uncertainty of the IMF does not affect the rates by more than a factor of $\sim 2$. Any change
    to the IMF slope for massive stars requires a change of SFR in a way that counteracts
    the impact of IMF variations on compact object merger rates.
    In contrast, we suggest that the uncertainty in cosmic SFR
    at low metallicity can be a significant factor at play. 
   }
  
   \maketitle
   
\section{Introduction}
The Laser Interferometer Gravitational-wave Observatory (LIGO) began its 
first upgraded observational run (O1) in September 2015; the first ever detection of a gravitational wave signal from 
a binary black hole (BH-BH) coalescence came 
shortly afterwards: i.e. GW150914 \citep{LIGO2016_first,LIGO2016_GW150914}.
Since then, four additional BH-BH mergers and, most recently, a double neutron star 
(NS-NS) merger, were detected and reported to the community: i.e.
GW151226 \citep[BH-BH,][]{LIGO2016_GW151226}, 
GW170104 \citep[BH-BH,][]{LIGO_GW170104}, 
GW170608 \citep[BH-BH,][]{LIGO2017_GW170608},
GW170814 \citep[BH-BH,][]{LIGO2017_GW170814}, and
GW170817 \citep[NS-NS,][]{NSNS_discovery_paper}.
The last three events were observed during the second observational run (O2); the Advanced Virgo detector \citep{AdvancedVirgo} joined the run on August 1, 2017
and contributed to the analysis of GW170814 and GW170817.

The LIGO discovery marks the beginning of the gravitational-wave era. 
Detections of the coalescence signals 
from compact binary mergers (CBM) are of utmost astrophysical significance as,
among other applications, they will constrain
potential formation scenarios, stellar
evolution models, and other assumptions associated with theoretical predictions
\citep[e.g.][]{Stevenson2015,Richard2017,Barrett2018}.
Various formation scenarios for CBM have been
proposed. Those most widely discussed include the isolated binary
evolution channel involving a common envelope (CE) phase
\citep[][]{Eldridge2016,B16Nat,Chruslinska2018,Mapelli2017,Giacobbo2018,Stevenson2017_compas}
or stable mass transfer \citep{vandenHeuvel2017},
isolated evolution of field triples \citep[eg.][]{Antonini2016},
dynamical evolution in dense stellar environments such as globular
clusters \citep[GCs;][]{Rodriguez2016a,Rodriguez2016b,Askar2017,Park2017}, 
nuclear star clusters \citep[][]{Miller2009,Antonini2016} or even
discs of active galactic nuclei \citep{Stone2017}),
and the formation of compact objects in very close and tidally locked
binaries through chemically homogeneous evolution 
\citep{deMink2016,Mandel2016,Marchant2016}.
We note that while  it is still possible to distinguish between different types of mergers
(i.e. BH-BH, BH-NS, or NS-NS) in a model-independent way \citep{Mandel2015,Mandel2017} using their gravitational-wave signatures, we still
lack a reliable way to determine the formation channel of 
an observed merger.
This is especially true in case of the LIGO/Virgo NS-NS merger \citep{Belczynski_NSNS_origin}.
In the case of BH-BH mergers there is some hope connected
to the measurement of the BH-BH spin-orbit misalignments \citep{Stevenson2017_compas,Farr2017,FarrHolzFarr2018}, 
although this may not work if the BH spins are intrinsically very small \citep{Belczynski2017_GW170104}.

Regardless of the formation scenario, the theoretical predictions for the
compact binary merger rates are burdened with significant uncertainties due to numerous 
assumptions and models with poorly constrained parameters, for example the infamous CE phase \citep{Dominik2012} or
the BH/NS natal kicks \citep{Repetto2015,Belczynski2016_kicks}).
One of the key ingredients in the calculations are the initial conditions for the simulations:
the birth properties of stellar clusters in the case of dynamical scenarios 
and the characteristics of primordial binaries in the case of isolated binary evolution channels.

Recently, \citet{dMB15} incorporated the updated primordial binary
parameter distributions obtained by \citet{Sana2012} from spectroscopic
measurements of massive O-type stars in very young ($\sim$ 2 Myr) open star
clusters and associations. The updated distributions included a much stronger
bias towards close binary orbits with respect to the previously adopted {\"O}pik's law,
i.e. a flat in logarithm distribution \citep{Opik1924,Abt1983}. 
Intuitively, this change should favour interacting binaries (including those undergoing 
the CE phase) and possibly cause a notable increase in merger rates. 
However, \citet{dMB15} found only a very small increase of less than a factor of 2. 
This is because the distributions
obtained by \citet{Sana2012} also show a heavy bias towards low eccentricities 
with respect to the thermal-equilibrium distribution of \citet{Heggie1975};
this distribution results in a nearly unchanged distribution of periastron 
separations, which is the essential separation regulating the onset of mass transfer. 

The sample of binaries observed by \citet{Sana2012} suffers from a significant
limitation: it is restricted to systems with $\logpday$ $<$ 3.5 (spectroscopically
detectable binaries) and dominated by very short-period orbits 
with P $<$ 20 days (hence the huge fraction of circularized systems). Since
the BH-BH mergers can originate from primordial binaries of up to 
log P $\sim$ 5.5 \citep{dMB15} the binary parameter distributions obtained by
\citet{Sana2012} need to be extrapolated to longer periods, which automatically
assumes no intrinsic correlations between parameters. However, the joint
probability density function cannot
be decomposed into independent distribution functions of the individual parameters, i.e.
$f(M_1, q, P, e) \ne f(M_1)f(q)f(P)f(e)$.
Observational studies have hinted at probable correlations \citep{Abt1990,Duchene2013},
but hitherto the selection biases have been too large to accurately quantify the
intrinsic interrelations. Recently, \citet[][hereafter \citetalias{md16}]{md16}
analysed more than 20 surveys of massive binary stars, corrected for their
respective selection effects, combined the data in a homogeneous manner, and fit
analytic functions to the corrected distributions. These authors confirm that many of the
physical binary star parameters are indeed correlated at a statistically
significant level.

\label{sec:var_with_Z}
\citet{dMB15} concluded that the most significant variations of merger
rates associated with the initial parameters are due to uncertainties of the initial
mass function (IMF) power-law slope for massive stars (merger rates going up and down by a factor
of six in the case of BH-BH). Cosmological calculations of the BH-BH merger rates
\citep[e.g.][]{Dominik2013,Dominik2015,B16Nat,Kruckow2018,Mapelli2017} are performed based on the
assumption of a universal IMF across the cosmic time. 
Often assumed is a so-called ``canonical'' IMF, which is a 
multi-part power-law distribution $dN/dM = \xi(M) \propto M^{-\alpha_i}$,
where $\alpha_1 = 1.3$ for $M/M_{\odot} \in [0.08,0.5]$ and $\alpha_2 = \alpha_3 = 2.3$ for
$M/M_{\odot} \in [0.5,1.0]$ and $[1.0,150.0]$, respectively \citep{Kroupa2001}.

Although clear evidence for strong IMF variations with environmental conditions is still lacking, 
there are a growing number of results hinting at departures from the IMF universality \citep[see reviews by][]{Bastian2010,Kroupa2013}.
Notably, a recent spectroscopic survey of massive stars in the 30~Doradus star-forming region in the Large Magellanic Cloud has
led to a discovery of an excess of stars with masses above $30 \msun$ with respect to the canonical IMF \citep{Schneider2018}; the best-fitted single power-law exponent for $M > 1 \msun$ is $\alpha_3 = 1.90_{-0.26}^{+0.37}$ \citep[although see a technical comment 
on the data analysis from][suggesting somewhat 
larger values for $\alpha_3$]{Farr2018}.

The unknowns associated with the massive-star IMF are often considered to be one of the significant contributors to uncertainty of 
compact binary merger rates calculated based on population synthesis.
In this work, we argue that by normalizing the simulated stellar population to the total amount of far-UV light that it emits, rather 
than to its total mass, one can significantly reduce the uncertainty associated with possible  
variations of the IMF slope in different environments; see Sect.~\ref{sec:small_imp}. 
As an example of such a variation and its impact on merger rate calculations,
we study in detail the case of a possible correlation between the massive-star IMF slope and metallicity Z.

With the exception of above-mentioned results of \citet{Schneider2018},
numerous observations of OB associations and clusters in the Local Group 
did not reveal any significant deviations from the canonical IMF slope for massive stars $\alpha_3 \approx 2.3$ \citep{Massey2003}.
These included surveys of the Milky Way \citep[Z $\approx \zsun$ = 0.02; ][]{Daflon2004} \footnote{throughout this study, 
we adopt $\zsun$ = 0.02 \citep{Villante2014}}, the Small Magellanic Cloud \citep[Z $\approx$ 0.004; ][]{Korn2000}
and the Large Magellanic Cloud \citep[Z $\approx$ 0.008; ][]{Korn2000},
indicating that for Z $\geq$ 0.004 ([Fe/H] $\gtrsim -0.7$) the high-mass end of IMF does
not significantly depend on metallicity. \citet{Moe2013} showed that the same is true
for the parameters of close binaries with massive stars. 
Even at solar metallicity, the discs of massive protostars are already highly prone to gravitational instability 
and fragmentation, explaining why the close binary fraction of massive stars is so large \citep{Kratter2006,Kratter2008,Kratter2010,md16}.
Further reducing the metallicity can therefore only marginally increase the close binary fraction of 
massive stars \citep{Tanaka2014,Moe2018}.
However, as we show,
the vast majority of BH-BH mergers evolving from the CE channel are expected to originate
from Z $<$ 0.004 environments, for which there is no direct observational evidence for
the persistence of the canonical IMF.

From a theoretical point of view, both the Jeans-mass formalism
\citep{Jeans1902,Larson1998,Bate2005,Bonnell2006} and the model of stellar formation
as a self-regulated balance between the rate of accretion from the proto-stellar
envelope and the radiative feedback from the forming star \citep{Adams1996} predict
that at a certain sufficiently low metallicity the IMF becomes top heavy
\footnote{ an IMF shifted towards higher stellar masses with respect to the canonical IMF, i.e. $\alpha_3 < 2.3$}.
In the first case, the prediction comes from the fact that the Jeans mass has to be
larger at lower Z owing to less effective cooling of the proto-stellar cloud
($M_J \propto \rho^{-1/2} T^{3/2}$). The radiative feedback, on the other hand, is also
metallicity dependent since photons couple less effectively to gas of lower metallicity.
Hydrodynamical simulations of the formation of Population III stars ($\rm Z/Z_{\odot}$~$<$~10$^{-6}$)
demonstrate that the first generation of stars were almost exclusively massive OB-type main-sequence (MS) stars
\citep{Bromm1999,Yoshida2006,Clark2011}. However, by redshift $z$~$\approx$~10,
the mean metallicity of actively forming stars was $\rm Z/Z_{\odot}$~$\approx$~10$^{-3}$
\citep{Tornatore2007,Madau2014}.
For such Population II stars with intermediate metallicities,
the IMF is expected to be only moderately top heavy compared
to the canonical IMF 
\citep{Fang2004,Daigne2006,Greif2008}.

From the observational side, there is indirect evidence for top-heavy IMF
variations at cosmological times. The relative paucity of metal-poor G-dwarf stars in
the Milky Way \citep[the so-called G-dwarf problem; ][]{Pagel1989} can be explained
by applying an IMF, which is increasingly deficient in low-mass stars the earlier the
star formation took place \citep{Larson1998}. By modelling the abundances of Lyman-break
and submillimetre galaxies in the $\Lambda$CDM cosmology, \citet{Baugh2005} found that the
observations can be reproduced only if some episodes of star formation with a top-heavy
IMF were also present in addition to the canonical IMF. \citet{Nagashima2005} used the
same model with two modes of star formation to explain the elemental abundances in the
intracluster medium of galaxy clusters. \citet{Wilkins2008} pointed out that the local stellar mass density
is significantly lower than the value obtained from integrating the cosmic star formation
history with a Salpeter IMF \citep[single slope, $\alpha = 2.35$; ][]{Salpeter1955}. At
low redshifts (z $<$ 0.5) they manage to match the observations using a slightly top-heavy
($\alpha_3 = 2.15$) IMF. For higher redshifts, however, \citet{Wilkins2008}
argued that no universal IMF can reproduce the measured stellar mass densities.

The only quantitative calibration of the relation between the high-mass IMF and metallicity obtained
thus far has been made possible thanks to the observations of some GCs
in the Milky Way. \citet{Djorgovski1993} noticed that the higher metallicity GCs tend to
have a bottom-light present day mass function (i.e. a deficit of low-mass stars).
\citet{DeMarchi2007} later found that these GCs are also the least concentrated sources in
their observed sample. Such a trend contradicts basic cluster evolution theory -- GCs
are expected to be losing low-mass stars from their outer regions owing to their collapse into
dense, highly concentrated clusters. An explanation was put forth by \citet{Marks2008},
who proposed that low-mass stars could be unbound from mass-segregated GCs during expulsion
of their residual gas. This introduces a dependence on metallicity, since the process of gas
expulsion is expected to be enhanced in metal-rich environments \citep{Marks2010}. Finally,
\citet{Marks2012b} concluded that in order to provide enough radiative feedback to expel the
residual gas and match the characteristics of the observed GCs, their IMF had to be top heavy with the value of $\alpha_3$ decreasing with cluster metallicity.

It should be noted, however, that the model of residual gas expulsion applied by \citet{Marks2012b} 
relies on simplified assumptions concerning the radiative feedback. First, the amount of energy deposited 
from stars into the ISM is fitted to stellar models of only three different masses \citep[35, 60, and 85 $\msun$;][]{Baumgardt2008},
and second there is no metallicity dependence. 
Additionally, \citet{Marks2012b} assumed that the energy needed for residual gas removal is provided 
by the stellar winds and radiation only (e.g. no feedback from supernova taken into account) and that 
all the energy radiated by the stars is deposited into the ISM. 
Thus, the exact relation between the high-mass IMF slope and metallicity remains highly uncertain.  \newline
The purpose of this study is twofold. First, we aim to incorporate the interrelated initial binary parameter
distributions and multiplicity statistics obtained by \citetalias{md16} to determine their
effects on the predicted rates and properties of CBM. Second, we investigate 
the importance of possible IMF variations for the merger rate calculations using the
\citet{Marks2012b} calibration of the IMF dependency on metallicity as an example of such a variation.
To achieve this, we perform comparative population synthesis where we use the works of \citet{dMB15} and \citet{B16Nat}
as references (hereafter \citetalias{dMB15} and \citetalias{B16Nat}, respectively).
In Sect.~\ref{sec:init_distr} we describe the new initial conditions and compare these with
the previously used distributions. In Sect.~\ref{sec:comp_method} we describe our computational method. In Sect.~\ref{sec:results}
we compare the distributions of the initial parameters of double compact merger progenitors
in our simulations. In Sect.~\ref{sec:res_merger_rates} we present the impact of the incorporated changes on the
cosmological merger rates. In Sect.~\ref{sec:discussion}  we discuss the metallicity distribution of BH-BH mergers
in our simulations and the significance of the top-heavy IMF in 
low-Z environments for the LIGO predictions. We conclude in Sect.~\ref{sec:summary}.

\section{Initial distributions}
\label{sec:init_distr}

\subsection{Inter-correlated initial binary parameters}

In the present study, we account for intrinsic correlations between the initial binary star physical parameters.  
We used the distribution functions presented in section 9 of \citetalias{md16}.  The correlations in the MS binary initial conditions are thoroughly discussed
in that paper, but we summarize  in this section the main results pertinent to the formation of compact objects mergers. \citetalias{dMB15} 
found that the majority of compact object mergers derive from MS binaries with initial orbital periods $\logpday$ = 2$-$4. 
Although for our simulations we generated companions across all orbital periods according to the distribution functions provided in \citetalias{md16},
in this section we focus the discussion on the companion star properties across intermediate periods $\logpday$ = 2$-$4.

\begin{figure}
        \includegraphics[width=\columnwidth]{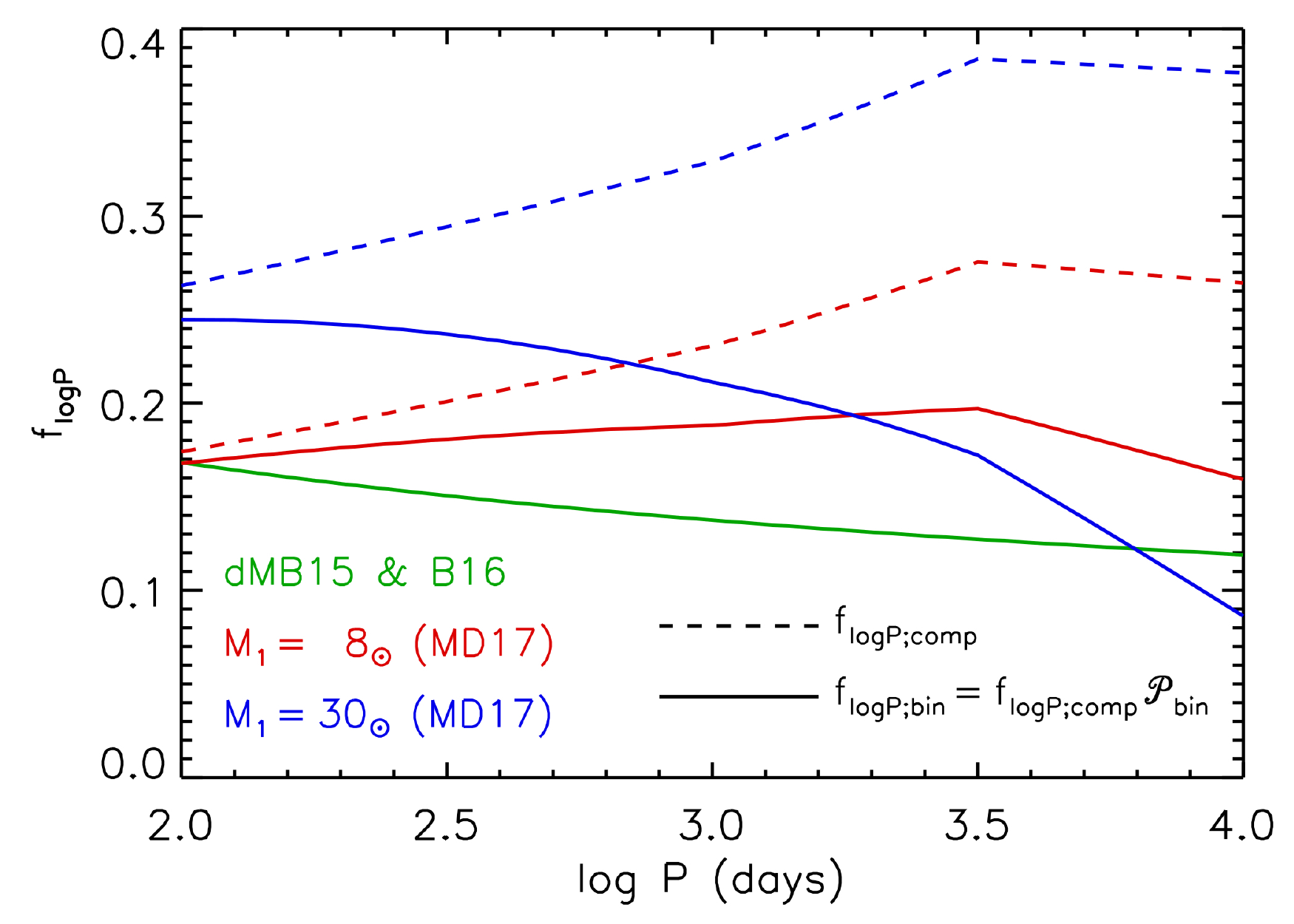}
    \caption{Frequency f$_{\rm logP}$ of companions with $q$ $>$ 0.1 per decade of orbital period across $\logpday$ = 2$-$4.  
    We compare the distribution f$_{\rm logP;comp}$ of all companions (binaries and tertiaries; dashed) to the distribution 
    f$_{\rm logP;bin}$ =  f$_{\rm logP;bin} \times {\cal P}_{\rm bin}$ of only the inner binaries (solid), where ${\cal P}_{\rm bin}$ 
    is the probability that the companion is a member of the inner binary.  The overall binary fraction and binary period distribution
    across $\logpday$ = 2$-$4 are similar between those adopted by \citetalias{dMB15} and \citetalias{B16Nat} (green) and the values for $M_1$ = $8 \msun$ (red)
    and $M_1$ = $30 \msun$ (blue) primaries based on the multiplicity statistics presented in \citetalias{md16}.
    }
    \label{fig:pdist}
\end{figure}

\textbf{Binaries versus triples:} \citetalias{dMB15} and \citetalias{B16Nat} assumed all companions to massive stars (M$>$10\Msun) were in binaries. In contrast,
\citetalias{md16} have modelled a companion star fraction and period distribution that includes true binaries as well as inner binaries
and outer tertiaries in hierarchical triples.  With increasing orbital period, the likelihood that a companion is 
an outer tertiary must increase.  In fact, essentially all wide companions ($\logp > 5$) to massive stars are the 
outer tertiaries in hierarchical triples/quadruples (\citealt{Sana2014}; \citetalias{md16}).  We model the probability that a 
companion is in the inner binary ${\cal P}_{\rm bin}(M_1,P)$ in a manner that reproduces the multiplicity
statistics derived in \citetalias{md16}.  Specifically, for $M_1$ = $8 \msun$, we find the probability decreases from 
${\cal P}_{\rm bin}$ = 0.96 for $\logp$ = 2 to ${\cal P}_{\rm bin}$ = 0.60 for $\logp$ = 4.  For $M_1$ = $30 \msun$,
the triple/quadruple star fraction is larger, and so we adopt ${\cal P}_{\rm bin}$ = 0.93 for $\logp$ = 2 and
${\cal P}_{\rm bin}$ = 0.23 for $\logp$ = 4.  We illustrate ${\cal P}_{\rm bin}(M_1,P)$ in Fig.~\ref{fig:pdist}.  

\textbf{Companion star fraction:}  \citetalias{dMB15} and \citetalias{B16Nat} implemented a binary star fraction such that 
F$_{\rm logP=2-4;bin} \approx$  28\% of massive primaries had binary star companions with periods $\logpday$ = 2$-$4 
and mass ratios $q$ = 0.1$-$1.0. The companion star fraction increases dramatically with primary mass
\citep{Abt1990,Raghavan2010,Sana2012,Chini2012,Duchene2013,Moe2013,md16}.
According to \citetalias{md16}, the intermediate-period companion star fraction increases from F$_{\rm logP=2-4;comp}$ = 0.46 
for $M_1$ = $8 \msun$ to F$_{\rm logP=2-4;comp}$ = 0.66 for $M_1$ = $30 \msun$. Approximately two-thirds of O-type primaries
have a companion with $\logpday$ = 2$-$4 and $q > 0.1$.  A significant fraction of these systems, especially those
with longer periods $\logpday$ = 3.5$-$4.0, are actually outer tertiaries in triples (see above and Fig.~\ref{fig:pdist}).

\textbf{Period distribution:} \citetalias{dMB15} and \citetalias{B16Nat} adopted a power-law binary period distribution $f_{\rm logP;bin}$ $\propto$ (log $P$)$^{-0.55}$
motivated by spectroscopic observations of massive binaries \citep{Sana2012}.  These works normalize this period distribution 
such that integration F$_{\rm logP=2-4;bin}$ = $\int_2^4 f_{\rm logP;bin} {\rm d(logP)}$ = 0.28 reproduces the binary fraction across
$\logpday$ = 2$-$4.  In the current study, we adopt a companion star period distribution $f_{\rm logP;comp}(M_1, P)$ based 
on the analytic fits in \citetalias{md16} that are shown in the lower panel of their Fig. 37.  We then derive the inner binary period 
distribution $f_{\rm logP;bin}$ =  $f_{\rm logP;comp} \times {\cal P}_{\rm bin}$.  In Fig.~\ref{fig:pdist}, we show $f_{\rm logP;bin}$ 
from \citetalias{dMB15} and \citetalias{B16Nat}, $f_{\rm logP;comp}$ for $M_1$ = $8 \msun$ and $30 \msun$ taken directly from \citetalias{md16}, and $f_{\rm logP;bin}$ for $M_1$ = $8 \msun$
and $30 \msun$ implemented in the current study.  While companions in general are weighted towards longer periods, the inner binary
period distribution is skewed towards shorter periods as implemented in \citetalias{dMB15} and \citetalias{B16Nat} and found in \citet{Sana2012}.

\begin{figure}
        \includegraphics[width=\columnwidth]{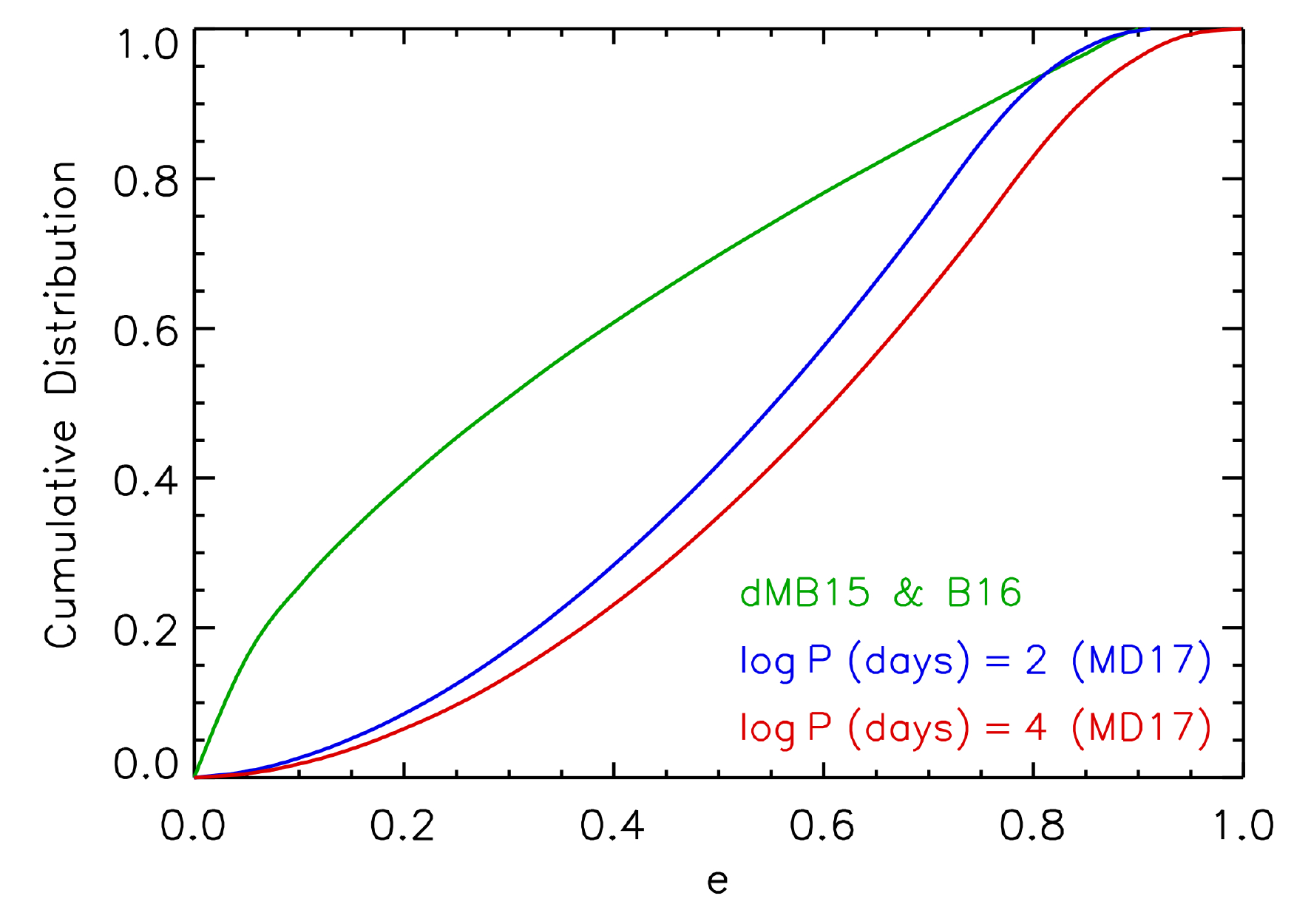}
    \caption{Cumulative distribution function of eccentricities.  \citetalias{dMB15} and \citetalias{B16Nat} adopted an eccentricity distribution $p_e$ $\propto$
    $e^{-0.4}$ (green) based on a sample of massive spectroscopic binaries \citep{Sana2012} that are dominated by short-period
    systems P $<$ 20 days and therefore weighted towards lower eccentricities $<$\,$e$\,$>$ = 0.3.  \citetalias{md16} (and references therein)
    found that massive binaries with intermediate (blue) and long (red) orbital periods are weighted towards higher eccentricities 
    $<$\,$e$\,$>$ = 0.6 and are sufficiently modelled by a power-law distribution $p_e$ $\propto$ $e^{0.8}$ with a turnover above $e$ $>$ 0.8.
    }
    \label{fig:edist}
\end{figure}

\textbf{Eccentricity distribution:} \citetalias{dMB15} and \citetalias{B16Nat} incorporated a power-law eccentricity distribution $p_e$ $\propto$ $e^{\eta}$ 
with exponent $\eta$ = $-$0.4 across the domain $e$ = 0.0$-$0.9 , as motivated by \citet{Sana2012}.
However, the \citet{Sana2012} sample of spectroscopic binaries are dominated by systems with P $<$ 20 days,
therefore the power-law slope $\eta$ = $-$0.4 is only appropriate for such short-period systems. \citetalias{md16} have found that eccentricities become 
weighted towards higher values with increasing orbital period, which they model with two parameters. First, \citetalias{md16} have defined the domain 
of the eccentricity distribution across the interval $e$ = 0.0--$e_{\rm max}$, where $e_{\rm max} (P)$ is the maximum eccentricity 
possible without substantially filling the Roche lobes of the binary (Eqn. 3 in \citetalias{md16}).  For P = 5 days, the binaries must have 
$e$ $<$ $e_{\rm max}$ $\approx$ 0.5 to be initially non-Roche-lobe filling.  Meanwhile, binaries with $\logpday$ = 2$-$4 can extend
up to $e_{\rm max}$ $\approx$ 0.95. Second, \citetalias{md16} have found the power-law slope $\eta$ also increases with increasing period.  For massive 
binaries, they fit $\eta$ $\approx$ 0.8 for $\logpday$ = 2$-$4.  We adopt the power-law slopes $\eta(M_1,P)$ presented in MD17 across the 
domain $e$ = 0$-$0.8$e_{\rm max}$.  At high eccentricities $e$ = (0.8$-$1.0)$e_{\rm max}$, we assume the probability distribution function
turns over according to a decreasing linear function such that $p_e$($e$ = $e_{\rm max}$) = 0. In Fig.~\ref{fig:edist}, we compare the cumulative 
distribution function of the eccentricity distribution adopted in \citetalias{dMB15} and \citetalias{B16Nat} to the updated distributions at $\logpday$ = 2 and 4.  
As expected, the eccentricity distribution based on the \citet{Sana2012} sample is skewed towards lower values.

\begin{figure}
        \includegraphics[width=\columnwidth]{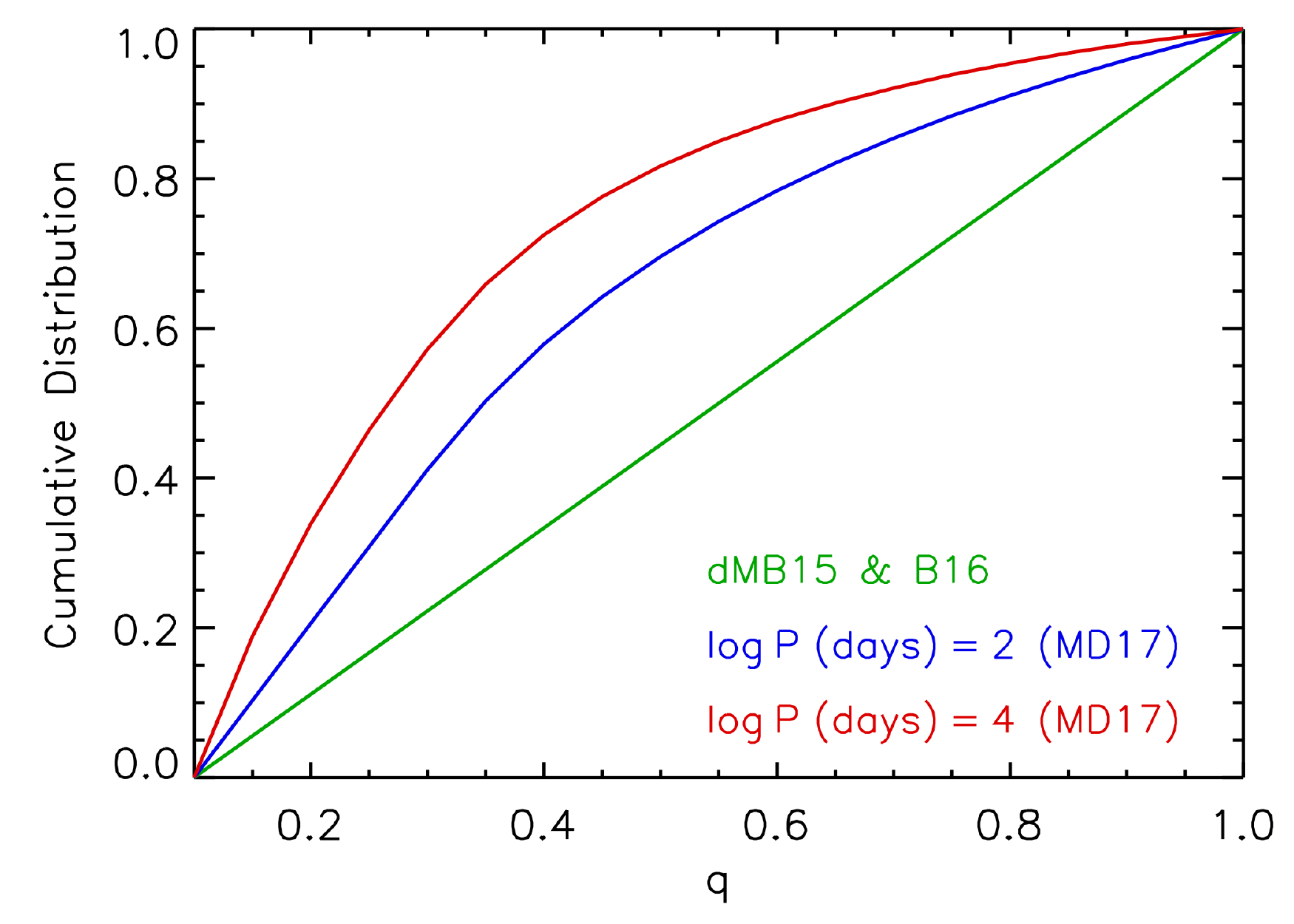}
    \caption{Cumulative distribution function of mass ratios.  \citetalias{dMB15} and \citetalias{B16Nat} adopted a uniform mass-ratio distribution $p_q$ $\propto$
    $q^0$ (green) based on a sample of massive spectroscopic binaries \citep{Sana2012} that are dominated by short-period systems P $<$ 20 days.
    \citetalias{md16} (and references therein) found that massive binaries with intermediate periods (blue and red) are weighted significantly towards 
    smaller mass ratios.
    }
    \label{fig:qdist}
\end{figure}

\textbf{Mass ratio distribution:} \citetalias{dMB15} and \citetalias{B16Nat} adopted a uniform mass-ratio distribution $f_q$ = $q^{0}$.  Again, this was based 
on a sample of massive spectroscopic binaries \citep{Sana2012} that is dominated by short-period systems $P$ $<$ 20 days. 
Based on a series of observational evidence (long baseline interferometry, companions to Cepheids, eclipsing binaries, adaptive optics,
Hubble imaging, etc.), \citetalias{md16} have demonstrated that slightly wider massive binaries are weighted considerably towards smaller mass ratios. 
Across intermediate periods $\logpday$ = 2$-$4, \citetalias{md16} have found the mass-ratio distribution is accurately described by a two-component
power-law $p_q$ $\propto$ $q^{\gamma}$ with slopes $\gamma_{\rm smallq}$ across small mass ratios $q$ = 0.1$-$0.3 and $\gamma_{\rm largeq}$ 
across large mass ratios $q$ = 0.3$-$1.0.  For massive binaries and $\logpday$ = 2, \citetalias{md16} have fit $\gamma_{\rm smallq}$ = 0.0 and 
$\gamma_{\rm largeq}$ = $-$1.4, while for $\logpday$ = 4, they have measured $\gamma_{\rm smallq}$ = $-$0.7 and $\gamma_{\rm largeq}$ = $-$2.0 (see their Fig.~35). 
In Fig.~\ref{fig:qdist}, we compare the cumulative mass-ratio distributions used by \citetalias{dMB15} and 
\citetalias{B16Nat} to the updated distributions measured at $\logpday$ = 2$-$4.

\textbf{Very massive binaries}: At present, there are no reliable observational measurements of the properties of very massive binaries with
intermediate periods $\logpday$ = 2$-$4 and primary masses $M_1$ $>$ $40 \msun$.  At these intermediate separations, the mass-ratio 
and eccentricity distributions do not significantly change across primary masses $8 \msun$ $<$ $M_1$ $<$ $40 \msun$.  In addition, the inner binary 
fraction and inner binary period distribution across $\logpday$ = 2$-$4 varies only slightly across the primary mass interval $8 \msun$
$<$ $M_1$ $<$ $40 \msun$ (see Fig.~\ref{fig:pdist}).  We therefore assume that systems with $M_1$ $>$ $40 \msun$ have the same period, eccentricity, 
mass-ratio distributions, and binary fractions as those systems with $M_1$ = $40 \msun$.  

\textbf{Triple-star evolution:} \label{sec:triple_evolution}
We model the evolution of all inner binaries with StarTrack. For triples, we ignore the outer tertiaries except 
for the following case.  If the inner binary initially has P $<$ 5 days and q $<$ 0.4, it most likely will undergo unstable Case A Roche lobe overflow (RLOF), 
merge on the MS, and effectively evolve as a single star.  If the tertiary has log $P_{\rm outer}$ (days) 
$\lesssim$ 4, then such a triple may lead to the formation of compact object merger.  We model these triples with inner binaries 
$P_{\rm bin}$ $<$ 5 days and $q_{\rm bin}$ $<$ 0.4 as simple binaries, where the original inner binary merges to become the effective 
primary with combined mass $M_1$ + $M_2$ and the tertiary becomes the effective binary companion with mass $M_3$. 
For simplicity, we assume no rejuvenation of the merger product, i.e. evolve it as a star with the same age as its companion 
with mass $M_3$. About 10\% of massive inner
binaries with $P$ $<$ 5 days have outer tertiaries in tight, hierarchical configurations log $P_{\rm outer}$ (days) $<$ 4 \citepalias{md16}.  
This triple-star channel certainly does not dominate, but can contribute an additional $\approx$5--10\% to the overall compact merger
rate. 

\subsection{IMF and its dependence on metallicity}
\label{sec:imf_z_dependance_introduced}

We adopt the same  IMF that was used by \citetalias{B16Nat}, 
as guided by the recent observations \citep{Bastian2010},
\begin{equation}
\label{eq:IMF}
 dN/dM = \xi(M) \propto  
 \begin{cases}  
      M^{-1.3}, & \text{for }  M/M_{\odot} \in [0.08,0.5]\\
      M^{-2.2}, & \text{for }  M/M_{\odot} \in [0.5,1.0]  \\
      M^{-\alpha_3}, & \text{for }  M/M_{\odot} \in [1.0,150.0]
 \end{cases} 
,\end{equation}
where for our standard model $\alpha_3$ = $2.3;$ we note that \citetalias{dMB15} adopt $\alpha_3$ = $2.7$ but also investigate $\alpha_3$ = $2.2$ and $3.2$).

Additionally we calculate a model in which the IMF slope for massive stars depends on metallicity, as described below.
In the following, we assume a conversion between [Fe/H] and Z given by \citet{Bertelli1994}, 
appropriate for stars with solar-like fraction of iron among all the metal components.

Based on the observations of stellar clusters in the Milky Way and N-body simulations 
(Sect.~\ref{sec:var_with_Z}) \citet{Marks2012b} calibrated that
$\alpha_{3;{\rm Marks}} = 2.63 + 0.66$[Fe/H] for [Fe/H] $<$ $-0.5$, which corresponds to Z $\lesssim$ 0.0063. 
We note that the above relation comes from a rather uncertain fit \citep[see Fig.~4 of][]{Marks2012b} and is based on a model with 
several caveats (see discussion at the end of Sect.~\ref{sec:var_with_Z}).
Because the observations of OB associations and clusters in the Local Group do not support any significant deviations from $\alpha$ = 2.3
for metallicities Z $\geq$ 0.004 \citep{Massey2003}, we modify the relation for $\alpha_3$ proposed by \citet{Marks2012b} 
such that the IMF dependence towards lower metallicity
does not occur until $\rm Z < 0.004$, and $\alpha_3$ = 2.3 for $\rm Z \geq 0.004$. 
\citet{Marks2012b} observational data extends down to about Z = 0.0001 ([Fe/H] $\approx$ $-$2.3, see their Fig.~4), 
and we decided to limit the further decrease of $\alpha_3$ for lower metallicities. 
Explicitly we adopt the following formula:
\begin{equation}
\label{eq:IMF_Z}
 \alpha_3({\rm Z}) = 
 \begin{cases}  
      2.3, & \text{for }  Z \geq 0.004\\
      2.3 + 0.76 \, [{\rm log\;Z}  + 2.4], & \text{for }  0.0001 < Z < 0.004 \\
      1.1, & \text{for }  Z < 0.0001 
 \end{cases} 
.\end{equation}
For some example values of Z, this formula yields $\alpha_3$(Z\,=\,$0.1 \zsun$)\,=\,2.07 and  $\alpha_3$(Z\,=\,$0.01 \zsun$)\,=\,1.31.

We note that we treat the IMF d$N/$d$M$ (all stars) and primary mass function d$N/$d$M_1$ (single stars and primaries in binaries) interchangeably.
Across large masses $M$ $>$ $1\msun$, both the IMF and primary mass function have the same slope $\alpha_3$ = 2.3 \citep{Kroupa2013}. The IMF and primary mass function only differ slightly across small masses $M$ $<$ $1\msun$.
Since $\alpha_3$ (and its dependence on metallicity) is the most important parameter
for determining merger rates, the assumption that d$N/$d$M$ equals d$N/$d$M_1$ is justified.

\section{Computational method}
\label{sec:comp_method}

\subsection{Physical assumptions -- the StarTrack code}
\label{sec:physical_assumptions}
We utilized the \textsc{StarTrack} population synthesis code \citep{Belczynski2002,Belczynski2008,Dominik2012}.
\textsc{StarTrack} is developed based on analytical formulae for the evolution of a non-rotating star obtained by \citet{Hurley2000} from the fits to the 
grid of evolutionary tracks calculated by \citet{Pols1998}. 
The original \citet{Hurley2000} models were updated with the prescriptions for the wind mass loss from O-B type stars \citep{Vink2001}, Wolf-Rayet stars \citep{Hamann1998,Vink2005}, 
and enhanced mass loss rates for luminous blue variables \citep{Belczynski2010a}. For core-collapse supernovae we adopt the 
convection driven, neutrino enhanced ``rapid'' supernova engine 
from \citet{Fryer2012}, which reproduces the observed mass gap between NSs and BHs \citep{Belczynski2012}. The key parameter of this model,
dependent on the mass of the CO core at the time of explosion, is the fraction of material ultimately falling back onto the proto compact object ($f_b$) 
and contributing to the final remnant mass. The supernova kick velocity is drawn from a Maxwellian distribution with $\sigma_{1D}$ = 265\kms \citep{Hobbs2005}
and lowered proportionally to the amount of fallback, i.e. $V_{NK}$ = $V_{NK;Hobbs} (1-f_b)$. This effectively means that the most massive BHs
in our simulations (up to $\sim$\,15$\msun$ for $\zsun$ and $\sim$\,40$\msun$ for $0.1 \zsun$) 
are also typically those that receive a relatively small or zero natal kick (direct collapse).

We calculated the CE evolution in one step by applying the energy balance prescription of \citet{Webbink1984}. We adopt $\alpha_{\rm CE}$ = 1 for 
the efficiency of energy transfer. The envelope binding parameter $\lambda$ is taken from the fits of \citet{Xu2010} and depends on the radius and the
evolutionary state of the donor. There are large uncertainties concerning the stability of mass transfer initiated by a Hertzsprung gap (HG)
donor star. \citet{Pavlovskii2017} recently reported that for a large range of donor radii and masses such mass transfer may actually be stable, rather than 
leading to a CE phase. While revised criteria for a range of different metallicities and other parameters are still under preparation (Ivanova, private communication), we
opted to not follow the evolution of systems with HG donors in which our standard criteria for the mass transfer stability \citep[Eq.~49 of][]{Belczynski2008} 
indicate a dynamically unstable event.

Up until this point our physical assumptions are the same as in submodel B of \citetalias{dMB15} and standard model M1 of \citetalias{B16Nat}. The only recent update is the 
addition of mass loss due to pair-instability pulsations \citep[model M10; ][]{Belczynski2016_PIP}. Such pulsations can affect stars with massive helium cores $M_{\rm He}$ 
between $\sim$ 40$-$45$\msun$ and $\sim$ 60$-$65$\msun$
and deplete them of a significant amount of mass \citep[$M_{\rm ejecta} \sim$ 5$-$20$\msun$;][]{Heger2002}. We modelled this by assuming that 
stars with $M_{\rm He}$ = 45$-$65$\msun$ are subject to pair-instability pulsations and lose all their mass above 45$\msun$. We note that the inclusion of
pair-instability mass loss does not affect our predictions for detections of double compact mergers with the LIGO O2 observational run sensitivity \citep{Belczynski2016_PIP}, 
and is consistent with existing data \citep{Fishbach2017}.

\subsection{Cosmological calculations of the merger rates}
\label{sec:cosmo_rates_method}

We placed our simulated systems into a cosmological background by populating the Universe up to $z$ = 15.
We modelled binaries across 32 different metallicities 
covering the range from 0.005$\zsun$ to 1.5$\zsun$.
In our standard model (I1)
the IMF does not depend on metallicity, 
and so for each Z we use the same sample of $2 \times 10^6$ systems (single stars, binaries, triples) generated according to the multiplicity statistics
described in Sect.~\ref{sec:init_distr} and a primary mass function $\xi(M_1)$ $\propto$ $M_1^{-\alpha_3}$ with $\alpha_3$ = 2.3 across $5 \msun$ 
$<$ $M_1$ $<$ $150 \msun$. We only evolve binaries ($\sim$\,72\% of systems with $M_1$ $>$ 5$\msun$) and hierarchical triples ($\sim$\,3.6\%), in which the inner binary is 
likely to merge during an early case A mass transfer (Sect.~\ref{sec:triple_evolution}).
For submodel I2 we assume the metallicity-dependent primary mass function 
according to Eqn.~\ref{eq:IMF_Z}, and for each Z we generate a corresponding sample of $10^6$ initial systems.

We calculate the merger rate density as a function of redshift by integrating through the cosmic star formation rate (SFR) history
as described in see Sect. 4 of \citet{Dominik2015} and Sect. 2.2 of \citet{Belczynski2016_kicks}. 
For each redshift bin $\Delta z$ = 0.1 up to $z$ = 15 we use the cosmic SFR
\begin{equation}
{\rm SFR}(z) = 0.015 \frac{(1+z)^{2.7}}{1+[(1+z)/2.9]^{5.6}}~{\rm M}_{\odot} {\rm Mpc}^{-3} {\rm yr}^{-1}
\label{eq:SFR}
\end{equation}
\noindent from \citet{Madau2014} as implemented in \citetalias{B16Nat}. The contribution from each of the metallicities in our simulations 
at a given redshift of star formation $z_{\rm SF}$ is then calculated as a fraction of SFR($z_{\rm SF}$) according to
the mean metallicity evolution model from \citet{Madau2014}, 
increased by 0.5 dex to better fit observational data \citep{Vangioni2015}, i.e.
\begin{equation}
\log(Z_{\rm mean}(z))=0.5+\log \left(  {y \, (1-R) \over \rho_{\rm b}}
\int_z^{20} {97.8\times10^{10} \, {\rm SFR}(z') \over H_0 \, E(z') \, (1+z')} dz'
\right)
\label{eq:Zmean}
.\end{equation}
\noindent We adopt a return fraction $R=0.27$ (fraction of stellar mass returned into the 
interstellar medium), a net metal yield of $y=0.019$ (mass of metals ejected into 
the medium by stars per unit mass locked in stars), a baryon density 
$\rho_{\rm b}=2.77 \times 10^{11} \,\Omega_{\rm b}\,h_0^2\,\msun\,{\rm Mpc}^{-3}$ 
with $\Omega_{\rm b}=0.045$ and $h_0=0.7$, a SFR from Eqn.~\ref{eq:SFR}, and
$E(z)=\sqrt{\Omega_{\rm M}(1+z)^3+\Omega_{\rm k}(1+z)^2+\Omega_\Lambda)}$
in a flat cosmology with $\Omega_\Lambda=0.7$, $\Omega_{\rm M}=0.3$, $\Omega_{\rm k}=0$,
and $H_0=70.0\,{\rm km}\,{\rm s}^{-1}\,{\rm Mpc}^{-1}$.
We assume a log--normal distribution of metallicity with $\sigma=0.5$ dex around the mean value at each 
redshift \citep{Dvorkin2015}. 

\citet{Madau2014} obtained their cosmic SFR based on the far-UV (FUV; 1500 $\AA$) and IR (8$-$1000 $\mu {\rm m}$) luminosity functions.
Both are regarded as good tracers of star formation as they are dominated by the contribution 
from short-lived massive stars: FUV directly and IR through re-radiation of dust-absorbed UV. Specifically, 
\citet{Madau2014} converted strength of the 1500 $\AA$ line in the UV spectrum $L_{\nu}({\rm FUV})$ into 
a cosmic SFR by applying an adequate conversion factor ${\cal K}_{\rm FUV}$, such that 
${\rm SFR} = {\cal K}_{\rm FUV} \times L_{\nu}({\rm FUV})$ (their Eqn. 10).
To calculate ${\cal K}_{\rm FUV}$ they assume a Salpeter IMF with slope $\alpha$ = 2.35 across masses $M$ = 0.1$-$100$\msun$ \citep{Salpeter1955}.
It is therefore inconsistent to implement the above Eqn.~\ref{eq:SFR} in combination with a more
realistic Kroupa-like IMF that turns over below $M$ $\lesssim$ 0.5$\msun$.
We introduce a conversion factor ${\cal K}_{\rm IMF}$ for a given IMF such that
\begin{equation}
\label{eq:SFR_conversion}
 {\rm SFR}_{\rm IMF}(z) = {\cal K}_{\rm IMF} \times {\rm SFR}_{\rm Salpeter}(z)
,\end{equation}
\noindent where ${\rm SFR}_{\rm Salpeter}(z)$ is given by Eqn.~\ref{eq:SFR}.
In order to calculate ${\cal K}_{\rm IMF}$ for various IMFs 
we compute the corresponding UV spectra via \textsc{Starburst99} code, 
designed to model spectrophotometric properties of star-forming galaxies \citep{starburst1,starburst2,starburst3}. 
In other words, when changing the IMF, we normalize the SFR from \citet{Madau2014} in such a way
that the amount of UV observed at $\rm 1500 \AA$ stays the same. 
Details are given in Appendix~\ref{sec:app_conv_coeff}, 
together with a tabularized set of ${\cal K}_{\rm IMF}$ values at solar and subsolar metallicities 
for different Kroupa-like IMFs. 

In the model with metallicity-dependent IMF at each redshift of star formation $z_{\rm SF}$ 
we use the mean metallicity of that redshift $Z_{\rm mean}(z_{\rm SF})$ to calculate 
a high-mass IMF slope (Eqn.~\ref{eq:IMF_Z}). We then use an appropriate 
value of ${\cal K}_{\rm IMF}$ from Table~\ref{tab:calib_coeff} 
to correct the cosmic SFR (Eqn.~\ref{eq:SFR_conversion}). For example, 
the conversion to Kroupa-like IMF \citep{Kroupa1993} with $\alpha_3$ = 2.3 \citep{Bastian2010}, 
the IMF which was assumed in a the recent studies \citep{B16Nat,Belczynski2016_PIP}, 
requires multiplying the cosmic SFR (Eqn.~\ref{eq:SFR}) by ${\cal K}_{\rm IMF; \alpha_3 = 2.3}$ = 0.51.
\footnote{Such a correction is missing in all our previous studies, except for \citet{Chruslinska2018}. However, its effect would be of secondary 
importance compared to uncertainties arising from evolutionary models (e.g. natal kicks), so their conclusions 
remain unchanged.}

The output of \textsc{StarTrack} and the cosmological calculations described above is in the form of CBM happening at different redshifts
. Each event is assigned with a normalization factor calculated as a
convolution of the cosmic SFR and the metallicity distribution at the redshift of the binary formation (Eqn. 7 of \citet{Belczynski2016_kicks}). Based on 
that we compute the source frame  merger rate density [$\rm Gpc^{-3} \, yr^{-1}$].

\subsection{Calculations of predictions for the LIGO/Virgo detectors}
\label{sec:det_rates_method}

Finally, 
we need to account for the detector sensitivity to produce predictions for the LIGO/Virgo observations. 
For each of our mergers we modelled the full inspiral-merger-ringdown waveform using the 
IMRPhenomD gravitational waveform templates \citep{Khan2016,Husa2016}.
We adopt a fiducial O2 noise curve (''mid-high'') from \citet{LVC2013}. 
Following \citet{B16Nat} we consider a merger to be detectable if the signal-to-noise
ratio in a single detector is above a threshold value S/N = $8$. 
We estimate detection rates as described in \citet{Belczynski2016_kicks}. 
This includes the calculation of detection probability 
$p_{\rm det}$, which takes into account the detector antenna pattern
\citep[LIGO samples a `peanut-shaped' volume rather than an entire spherical volume enclosed 
by the horizon redshift;][]{Chen2017}.
For increased accuracy with respect to \citet{Belczynski2016_kicks}, where we used an
analytic fit to obtain $p_{\rm det}$ \citep[Eqn. A2 of ][]{Dominik2015}, 
in this work we interpolate the numerical
data for the cumulative distribution function available on-line at
\url{http://www.phy.olemiss.edu/~berti/research.html}. This improvement
leads to a few per cent increase in detection rates. 

\subsection{Differences with respect to \citetalias{B16Nat} and \citet{Belczynski2016_PIP}}

\label{sec.differences_wrt_B16}

In terms of physical assumptions (Sect.~\ref{sec:physical_assumptions}),
in the current study, we adopt exactly the model M10 from \citet{Belczynski2016_PIP}. 
Model M1 from \citetalias{B16Nat} is also very similar, only differing by the lack 
of mass loss due to pair-instability pulsations.

There are, however, two differences between our current calculations and 
these two previous studies. We correct the assumed SFR (Eq.~\ref{eq:SFR}) 
derived from a Salpeter IMF for a more realistic Kroupa-like IMF (see Appendix~\ref{sec:app_conv_coeff}),
which lowers the predicted merger rates by a factor of $\sim$0.51 in the case of $\alpha_3$ = 2.3.
In both \citet{Belczynski2016_PIP}
and \citetalias{B16Nat} there was a mistake in the way the simulated systems
were normalized to match the entire stellar population (i.e. the calculation of the simulated mass 
$M_{\rm sim}$). As a result, the calculated merger rates were about $\sim$1.82 smaller. 
The net effect of these two changes is a slight decrease of the merger rates (by a factor of 0.926). 
This is why the numbers we present for model M10 in the following sections 
are 0.926 times smaller than the corresponding values presented in \citet{Belczynski2016_PIP}.
We note that in the most recent work, \citet{Belczynski2017_GW170104}, the correction 
factor of 0.926 is already applied to all the models. 

\section{Results}

\label{sec:results}

\subsection{Birth properties of compact binary mergers}

Starting from the zero-age MS (ZAMS), we evolved $\sim$\,$1.5 \times 10^6$ binaries 
with primaries of at least $5\msun$ drawn from the 
initial distributions of \citetalias{md16}.
According to the assumed IMF and multiplicity 
statistics, our sample constitutes about 29\% 
of the mass of all the stars forming at ZAMS, 
and so our simulations are normalized to a star-forming mass of
$M_{\rm sim} = 1.687 \times 10^{8}\msun$.
The formation efficiencies of CBM, 
depend significantly on metallicity, but are generally
between $10^{-5}$ and $10^{-7}$\,$[M_{\odot}^{-1}]$ (see Table~\ref{tab:ZAMS_mass_fractions}); we
define CMBs as the number of merging systems of a given type in our simulations
per unit of star-forming mass, i.e. $X_{\rm CBM} \; (\msun^{-1})= N_{\rm
CBM} / M_{\rm sim}$. 

We wcompare the birth properties of CBM progenitors
between the two simulated samples:
(1) one assuming the initial MS binary distributions of \citet[][old simulations from \citealt{Belczynski2016_PIP}]{Sana2012} 
and the other adopting the recent results of \citetalias{md16}.
We discuss the NS-NS mergers together with BH-BHs because these two merger types originate from systems with similar mass ratios 
and their distributions of initial pericentre separations and primary masses are clearly distinguishable.
The BH-NS progenitors are presented separately.

\begin{figure*}
      \includegraphics[width=\textwidth]{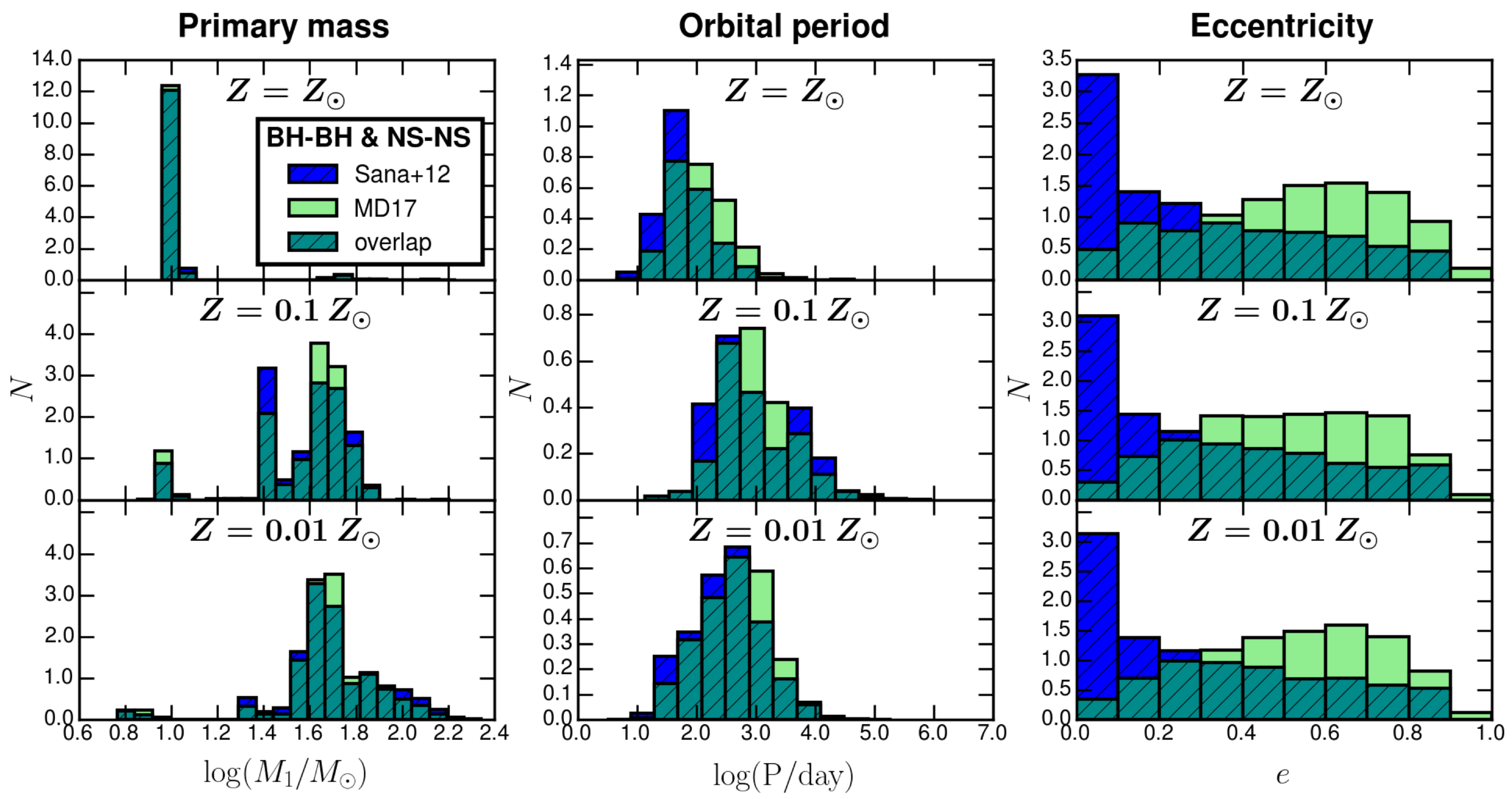}
      \includegraphics[width=\textwidth]{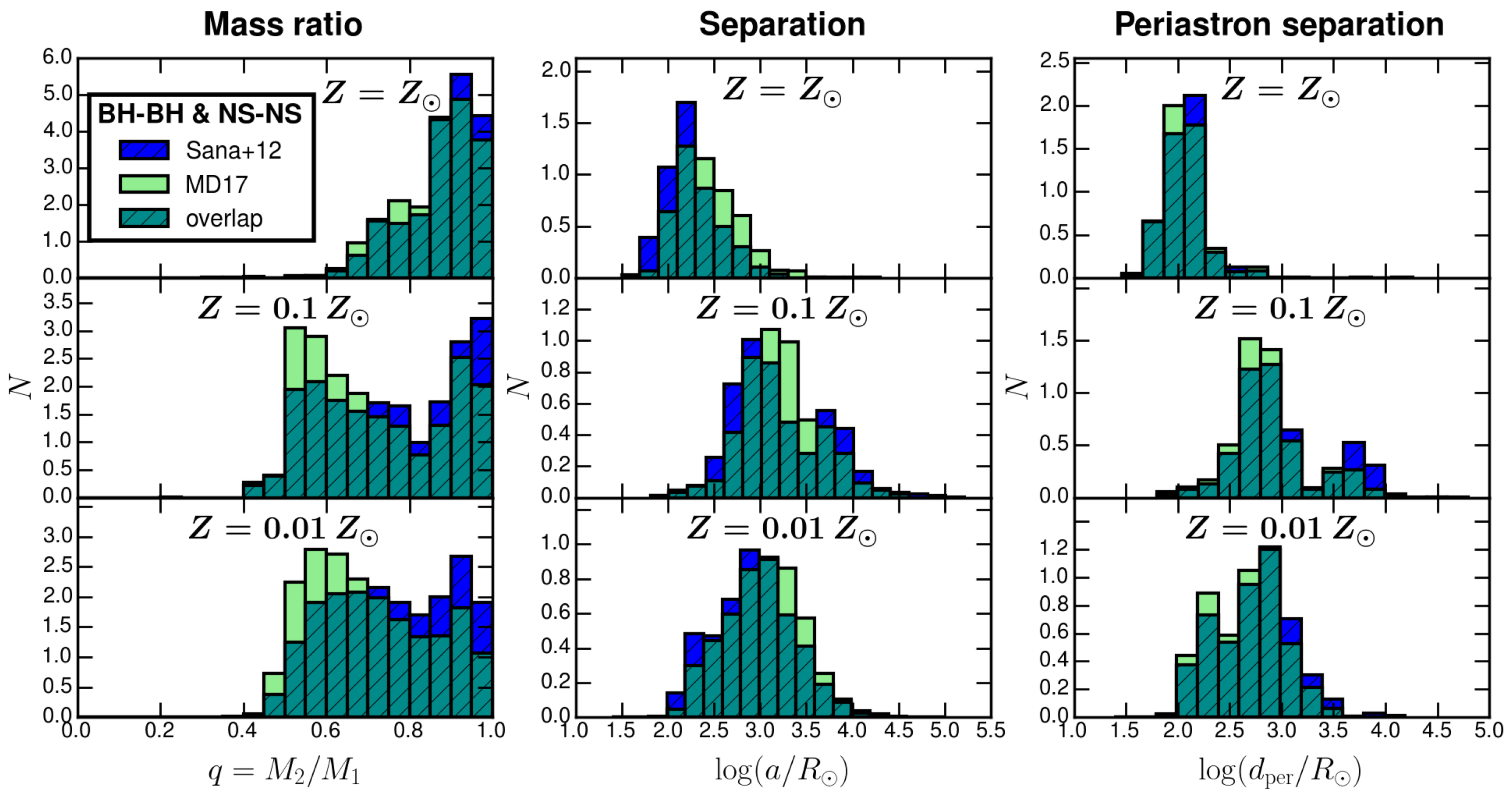}
    \caption{Birth distributions of initial ZAMS binary parameters of the progenitors of BH-BH and NS-NS mergers. The histograms 
    are normalized to unity for an easier comparison. NS-NS mergers dominate at $Z = \zsun$ (upper panel) while BH-BH mergers dominate at $Z = 0.1\zsun$ and $Z=0.01\zsun$ (lower panels).}
\label{fig.ZAMS_properties}
\end{figure*}

\begin{table}
\caption{Comparison of the formation efficiencies per unit of star-forming mass of 
CBM: BH-BH, BH-NS, and NS-NS between the two models with different initial distributions:
either \citet[][, old]{Sana2012} or \citetalias{md16}(new). The comparison is made at three different metallicities.
No metallicity-dependence for the IMF is assumed here. We indicate in bold text 
the dominant metallicity (out of the three showed here)
for the formation of each type of the double compact mergers.
}               
\centering          
\begin{tabular}{c| c c | c}     
\hline\hline       
  & \multicolumn{2}{c|}{Formation efficiency} &   \\ 
 Metallicity & \multicolumn{2}{c|}{$X_{\rm CBM}$ $(M_{\odot}^{-1})$\tablefootmark{a} } &  Relative \\ 
merger type\tablefootmark{b}   & Sana+12 & MD17 & MD17 / Sana \\ 
& (model M10) & (model I1) & \\
\hline          
   $Z = Z_{\odot}$ &&&\\
   NS-NS &  $\mathbf{8.3 \times 10^{-6}}$ &   $\mathbf{3.8 \times 10^{-6}}$ &  $0.46$ \\  
   BH-NS\tablefootmark{c} &  $9.0 \times 10^{-8}$ &   $1.1 \times 10^{-7}$ &  $1.22$ \\
   BH-BH &   $3.5 \times 10^{-7}$ &    $1.5 \times 10^{-7}$ &  $0.43$ \\
\hline                  
  $Z = 0.1\zsun$ &&&\\
   NS-NS &  $1.75 \times 10^{-6}$ &   $1.0 \times 10^{-6}$ &  $0.57$ \\  
   BH-NS &  $\mathbf{3.0 \times 10^{-6}}$ &   $\mathbf{1.2 \times 10^{-6}}$ &  $0.4$ \\
   BH-BH &   $2.1 \times 10^{-5}$ &    $8.4 \times 10^{-6}$ &  $0.4$ \\
\hline              
$Z = 0.01\zsun$ &&&\\
   NS-NS &  $1.7 \times 10^{-6}$ &  $1.2 \times 10^{-6}$ & $0.71$ \\  
   BH-NS &  $9.6 \times 10^{-7}$ &  $3.6 \times 10^{-7}$ & $0.38$ \\
   BH-BH &   $\mathbf{5.3 \times 10^{-5}}$ &  $\mathbf{2.9 \times 10^{-5}}$ & $0.55$ \\
\hline \hline
\end{tabular}
\tablefoot{\\
\tablefootmark{a}{Compact binary merger formation efficiency per unit of star-forming mass: $X_{\rm CBM} \; (\msun^{-1}) = N_{\rm CBM} / M_{\rm sim}$, 
where $M_{\rm sim;MD17} = 1.687 \times 10^{8}\msun$ and $M_{\rm sim;Sana} = 1.56 \times 10^{8}\msun$} \\
\tablefootmark{b}{All the systems expected to merge within the Hubble time.} \\
\tablefootmark{c}{Only a statistically insignificant (< 20) number of BH-NS mergers 
form at $Z=\zsun$ in our simulations.}
}
\label{tab:ZAMS_mass_fractions}      
\end{table}

\subsubsection{Progenitors of BH-BH and NS-NS mergers}

\label{sec:bhbh_nsns_prog}

In Fig.~\ref{fig.ZAMS_properties}, we show the birth properties of 
systems that evolve to form BH-BH and NS-NS mergers. We only include 
double compact binaries that are expected to merge within the Hubble time.
The histograms are normalized to unity for an easier comparison;
see Table~\ref{tab:ZAMS_mass_fractions} for the relative 
formation efficiencies.

\textbf{Primary masses}: The distributions of primary masses are noticeably 
different in different metallicities. The peak 
around $10 \msun$ at $Z = 0.02$ corresponds to NS-NS progenitors, which are the dominant 
type of CBM at solar metallicity. At $Z = 0.002$ the 
formation of BH-BH begins to dominate and happens for primary masses $M_1 > 20\msun$, 
peaking around $25$ and $45\msun$. This bimodal shape is a direct consequence of 
the bimodal distribution of CO core masses $M_{\rm CO}$ of supernova progenitors expected to undergo a direct
BH formation.
In the rapid supernova engine we adopted \citep{Fryer2012}, BHs receive no natal kick on formation if $M_{\rm CO}$ is either between $6$ and $7\msun$ or above $11\msun$ (see their Eqn.~16).
At $Z = 0.01\zsun$ the higher mass peak corresponding to $M_{\rm CO} > 11\msun$ dominates. 

The differences between the new and old initial distributions are very small. The NS-NS as well as 
lower mass BH-BH corresponding to $M_{\rm CO}$ = 6$-$7$\msun$ are most preferably formed from 
systems with very high initial mass ratios $q > 0.8$, which is why there is relatively less of such systems 
in the simulations with \citetalias{md16} initial conditions. 

\textbf{Orbital periods, eccentricities, and periastron separations}: The new initial binary period distribution 
of \citetalias{md16} is in general very similar in shape
to that proposed by \citet{Sana2012} in the crucial range of $\logpday$ between $2$ and $4$ (especially for systems with 
massive primaries $\sim$\,$30\msun$; see Fig.~\ref{fig:pdist}). However, there is a noticeable 
shift towards larger periods (and separations) among compact binary merger progenitors in the new simulations.
This can be explained in connection with the eccentricity distribution of these systems. In the case of \citetalias{md16} the eccentricity distribution is
slightly skewed towards $e > 0.5$, which is very different from that of \citet{Sana2012} showing a strong preference for nearly circular orbits.
As a result the periastron separation distributions in the new and old simulations are very similar.
This showcases the fact that the pericentre separation $d_{\rm per}$
 determines the evolutionary path of a binary with given component masses and decides whether or not it will 
form a compact binary merger.
In the evolutionary channel involving a CE phase, only a  fixed range of $d_{\rm per}$,
that is independent of the choice of initial properties, results in the formation of CBM.

\begin{figure*}
      \includegraphics[width=\textwidth]{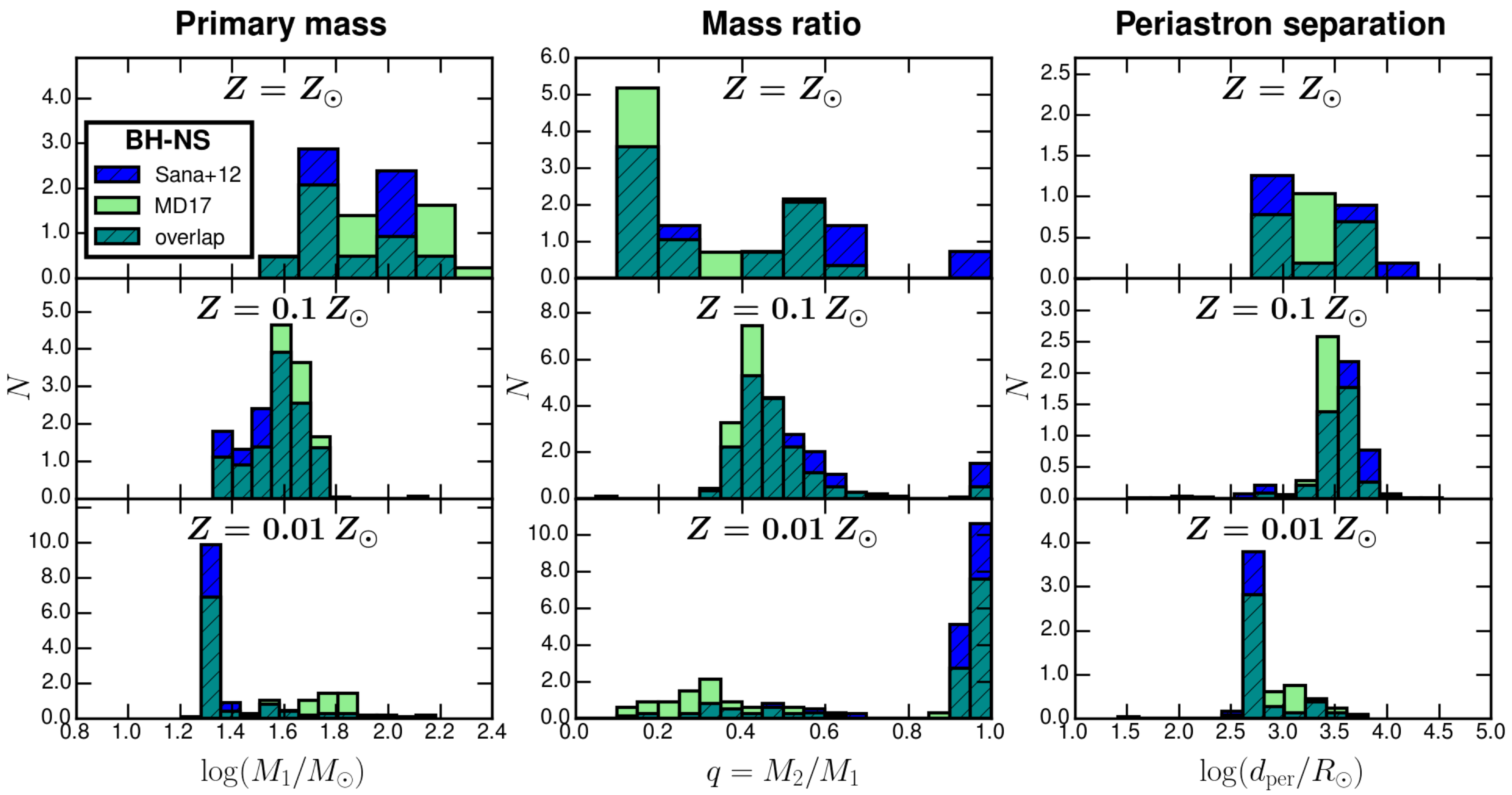}
    \caption{Birth distributions of initial ZAMS binary parameters of the progenitors of BH-NS mergers. The histograms 
    are normalized to unity for an easier comparison. We note that the upper panels ($Z=\zsun$) are made based on 
    a very small number of systems ($< 20$) that form in solar metallicity. }
\label{fig:BHNS_properties}
\end{figure*}

At $Z=\zsun$, where the formation of NS-NS dominates, the periastron separations are centred around $\sim$\,$100\rsun$. 
At $Z =0.1\zsun$ BH-BH become dominant and shift the $d_{\rm per}$ distribution towards higher values of over $\sim$\,$100\rsun$. At this 
metallicity the distribution is clearly bimodal with peaks around $\sim$\,$750\rsun$ and $\sim$\,$4000\rsun$. This comes from 
the fact that at $Z \sim 0.1 \zsun$ there are two relevant formation channels of merging BH-BH: the dominant channel ($\sim$70\% of the systems),
in which there is only one CE phase taking place
after the first BH has already formed (the peak at $d_{\rm per}$\,$\sim$\,$750\rsun$); and another channel ($\sim$20\% of the systems), in which there 
are two CE phases, one before each BH formation. The systems originating from the latter channel
are those responsible for the second peak at $d_{\rm per}$\,$\sim$\,$4000\rsun$. 
These systems need to have larger initial $d_{\rm per}$ to prevent any RLOF during the HG evolutionary stage of the primary;  we note that we do not 
allow for the CE phase from HG donors. Instead, the primary needs to initiate a RLOF and the CE phase later during the core helium burning phase. 
This channel becomes ineffective in our simulations 
for extremely low metallicities ($Z \lesssim 0.0005$), where stars do not increase their radii sufficiently after the HG stage.
For that reason at $Z=0.01\zsun$, the $d_{\rm per}$ distribution is no longer clearly bimodal. It is also slightly shifted towards smaller separations with respect 
to the distribution at $Z=0.1\zsun$, again because of smaller sizes of the stars. Overall, the BH-BH formation channel involving two different CE phases is only relevant (i.e. at least 
5\% of all the BH-BH formed this way) at metallicity range $Z$ \,=\,$0.0005$\,-\,$0.004$, contributing up to at most $\sim$\,22\% of the BH-BH 
formation at metallicity $Z$\,$\approx$\,$0.0025$.

We note that the secondary peak in pericentre separations around $\sim$\,$4000\rsun$ at $Z=0.1\zsun$ is noticeably smaller in the simulations 
with the new initial distributions.
This is because of the fact that, according to the multiplicity 
statistics of \citetalias{md16}, the number of inner binaries with massive primaries ($\gtrsim 30\msun$) having orbital
periods of about $\logpday \approx 3.5$ to $4.0$ is smaller with respect to the old distributions of \citet{Sana2012}; 
see Fig.~\ref{fig:pdist}.

\textbf{Mass ratios}: Merging NS-NS and BH-BH in our simulations originate almost exclusively from 
binaries with high mass ratios: $q_{\rm NS-NS} \gtrsim 0.6$ and $q_{\rm BH-BH} \gtrsim 0.5$, respectively 
(at $Z = 0.1\zsun$ the systems with initial $q<0.5$ are all BH-NS). The initial mass ratio distribution of \citet{Sana2012} is flat in $q$. Meanwhile, the 
new results of \citetalias{md16} include a distribution decreasing quickly with $q$ (see Fig.~\ref{fig:qdist}). This is reflected in the changes 
of the mass ratio distribution of CBM progenitors, which are slightly shifted towards $q$\,$\sim$\,0.5 at $Z = 0.1\zsun$ and $Z = 0.01\zsun$ 
in the new simulations. 

\subsubsection{Progenitors and formation of BH-NS mergers}
\label{sec:bhns_prog}

In Fig.~\ref{fig:BHNS_properties}, we show the birth properties of 
systems that evolve to form BH-NS mergers. 
This time, for simplicity, we focus only on the three most 
important parameters: primary mass, mass ratio, and periastron 
separation. 

The formation of BH-NS mergers is very different in the three different metallicites 
shown: $\zsun$, $0.1\zsun$, and $0.01\zsun$. In the solar metallicity there 
are hardly any BH-NS systems formed. The histograms for $Z=\zsun$ are based on only around a dozen binaries, 
and, as a result, these histograms are not very meaningful. The reason why the BH-NS formation 
is so ineffective in our simulations at solar metallicity is that the stars with $Z=\zsun$ already expand 
to their nearly maximum radius during the HG phase. This means that, in 
the majority of cases, the RLOF is initiated 
from the primary while it is a HG star. Because BH-NS systems typically originate from 
binaries with relatively small initial mass ratios of $q \lesssim 0.6$,
then, according to the standard \textsc{StarTrack} criteria,
such a mass transfer would most likely be flagged as unstable, leading to a CE phase
from a HG donor.
In this work, we decided to not follow the evolution of such systems 
(see Sect.~\ref{sec:physical_assumptions} for more details).

 \begin{figure*}
\sidecaption
  \includegraphics[width=12cm]{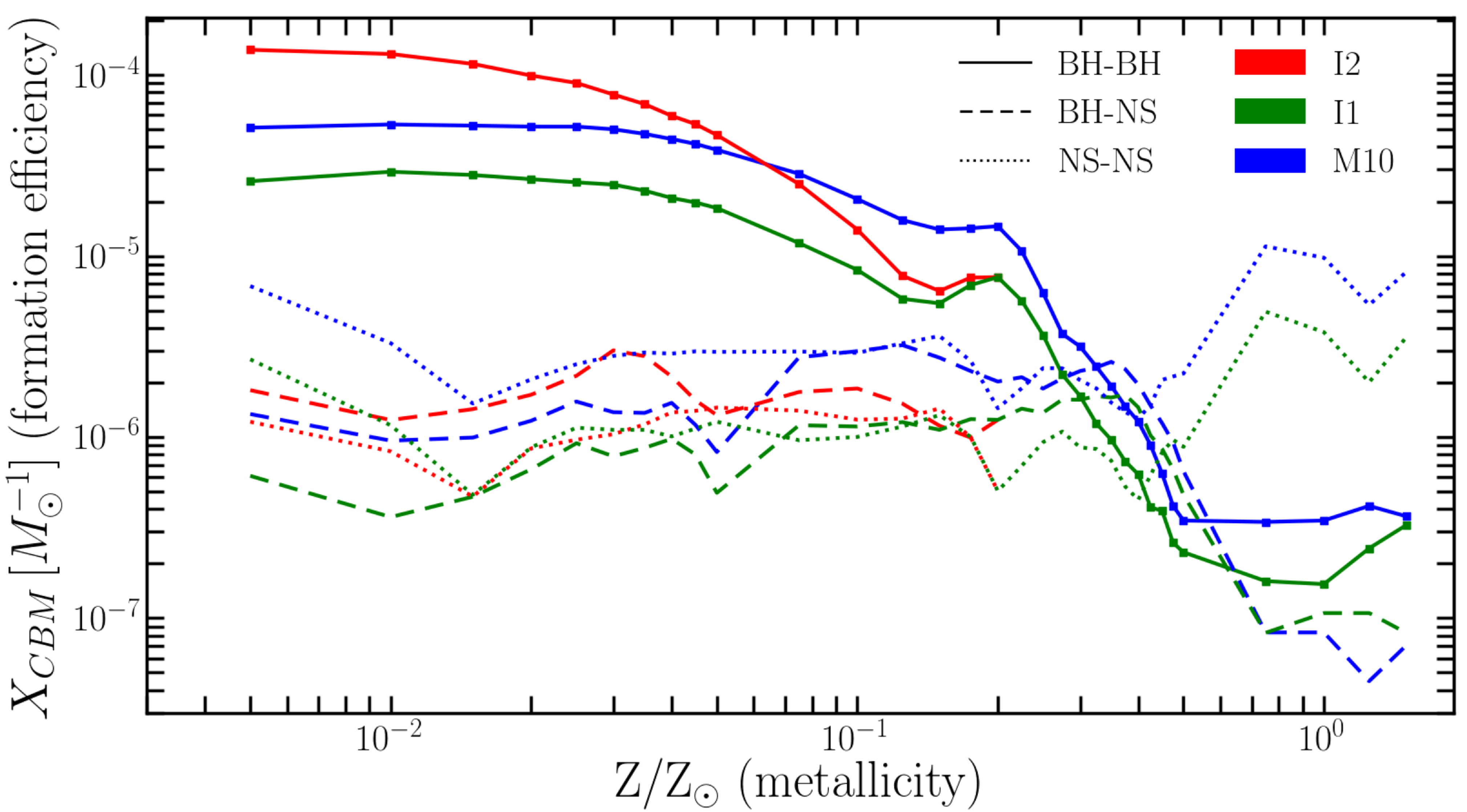}
     \caption{Formation efficiency (i.e. number of systems formed per unit of star-forming mass, $X_{\rm CBM} \; (\msun^{-1}) = N_{\rm CBM} / M_{\rm sim}$)
    of different types of CBM at different metallicities. We compute a grid of 32 different metallicities (indicated with 
    filled squares at the BH-BH curves), ranging from $0.005\zsun$ to $1.5\zsun$. We indicate three different models: M10 in blue (initial distributions 
    of \citealt{Sana2012}), I1 in green (initial distributions of \citetalias{md16}) and I2 in red (same 
    as I1 but with a top-heavy IMF at low metallicities; see Eq.~\ref{eq:IMF_Z}).
    We note that model I2 is only different from I1 at $Z<0.004 = 0.2\zsun$. }
     \label{fig:Xcmb}
\end{figure*}

At subsolar metallicity $Z=0.1\zsun$ the formation of BH-NS mergers 
is much more effective, and the initial distributions are representative for 
the majority of BH-NS mergers formed across all of the 32 metallicites we compute. 
The primary expands much more after the HG phase than 
in the case of solar metallicity
\citep[Fig. 2 of][]{Belczynski2010b}
and, as a core-helium burning giant with a convective envelope, initiates an unstable RLOF and a CE phase. 
After the BH formation, the secondary eventually evolves off the MS, 
expands, and initiates a second CE episode. As a result, the binary is very 
close, the natal kick velocity gained by a newly formed NS does not disrupt the system and the binary merges within the Hubble time. We note 
that this is the exact same evolutionary route (i.e. two CE phases)
that also operates in the case of BH-BH mergers in intermediate metallicites 
$Z$ \,=\,$0.0005$--$0.004$; see the paragraph on orbital periods in Section~\ref{sec:bhbh_nsns_prog}.

The distribution of primary masses at $Z=0.1\zsun$  is slightly shifted towards lower values 
(peak at $\sim40\msun$) with respect to the same distribution for BH-BH progenitors 
(peak at $\sim50\msun$). The dominant range of mass ratios 
is between $0.3$ and $0.6$ (the contribution from $q > 0.95$ is described below).
In the simulation with the updated initial distributions from \citetalias{md16}, the mass ratio distribution 
is slightly shifted towards smaller values. This, in turn, causes a small shift towards higher $M_1$ values, 
so that the secondary mass distribution (not explicitly shown here) stays roughly the same.
The optimal periastron separation is about 
$3000\rsun$ similar to the case of BH-BH mergers evolving through the 
same channel involving two CE phases;
see the right peak in the periastron separation distribution for 
$Z=0.1\zsun$ in Fig.~\ref{fig.ZAMS_properties}. We note that in the case of new 
simulations with the initial conditions of \citetalias{md16} there are fewer 
binaries with initial periods above $\logpday \approx 3.5$ with respect 
to the old distributions of \citet{Sana2012}, which is why the peak in the periastron 
separations is slightly shifted towards smaller values (also similarly to the BH-BH case at $Z=0.1\zsun$). 

In the case of very low metallicities (such as $Z = 0.01\zsun$) the main 
formation channel of BH-NS mergers described above is no longer efficient. This is 
because the lower the metallicity the more massive the BHs formed. For comparison, 
the least massive BHs in our simulations are of about $5.5\msun$ at $Z = 0.1\zsun$ 
and about $7.0\msun$ at metallicity $Z = 0.01\zsun$. As a result, at $Z = 0.01\zsun$, 
the mass ratio of the components at the point when the secondary expands and initiates a RLOF is not 
sufficiently high to cause an unstable mass transfer and a CE phase. 

At $Z = 0.01\zsun$ another formation channel dominates, which is also present, yet not so important
at higher metallicities. This channel involves binaries with a mass ratio close to unity ($q \gtrsim 0.9$),
comprised of stars with masses that are close to the threshold between the NS and BH formation
($M \sim 20\msun$). Because the binary components have similar masses, the secondary evolves 
transfer between the two core-helium burning stars, which reverses the mass ratio of the components.
The primary eventually collapses to form a BH in a direct event with no mass eject and no natal kick. The `Rapid' BH formation model of \cite{Fryer2012} predicts a 100\% mass fallback for the 
least massive BH progenitors with CO core masses between 6 and 7$\msun$; see their Eq.~16. 
The rejuvenated secondary continues to expand and initiates a second RLOF. This time the mass 
ratio is lower and the mass transfer becomes unstable, leading to a CE phase. The binary becomes 
close enough to survive the supernova and  NS formation, and later becomes a BH-NS merger.  
This scenario requires a specific range of initial binary parameters and is much less effective 
than the more standard evolutionary route dominating the BH-NS formation at intermediate metallicities.

\subsection{Formation of compact binary mergers across metallicity}
\label{sec:Xcbm_plot}

In Fig.~\ref{fig:Xcmb} we show the relation between the
formation efficiency $X_{\rm CBM} \; (\msun^{-1})$ and metellicity for different types of CBM. We note that similar trends were obtained recently by \citet{Giacobbo2018_new}.
It is evident that the formation of BH-BH mergers becomes 
most effective at low metallicities $Z \lesssim 0.1\zsun$.
This is a well-known result \citep[e.g.][]{Belczynski2010b,Dominik2012,Eldridge2016,B16Nat,Mapelli2017,Giacobbo2018_new}
for mainly two reasons: first, the fact that at lower metallicities the BH progenitors
are relatively more massive and, therefore, more BHs are formed through direct collapse with zero natal kick,
and second there is a higher chance for a CE event to be initiated by a core-helium burning 
donor, rather than by a HG donor, because of smaller radii of HG stars \citep[Fig.~2 of ][]{Belczynski2010b}.

According to stellar tracks from \citet{Hurley2000}, at metallicities close to the solar $Z \simeq 0.02$, 
there is only a small range of separations at which a mass transfer from core-helium burning donors
of $> 20 \msun$ is initiated. Much more likely, the giant causes a RLOF earlier during its 
rapid HG evolution, in which case we choose to not allow for a CE evolution. Apart from strong stellar winds, 
this is what suppresses the formation of BH-BH and BH-NS compact binaries at metallicities close to the solar 
$Z \simeq 0.02$ in our simulations.

The formation efficiency of BH-NS mergers $X_{\rm BHNS}\; (\msun^{-1})$ does not increase
in a similar fashion as $X_{\rm BHBH}\; (\msun^{-1})$ does at very low metallicities. 
There are two major factors at play. The main formation channel of BH-NS involving two CE phases 
becomes less effective in very low metallicities (Sect.~\ref{sec:bhns_prog}). Additionally, natal kicks for 
NSs formed in core collapse are significant even at very low metallicities.

The formation of NS-NS mergers is much less metallicity-dependent, as was similarly found by \citet{Giacobbo2018_new}.
It is slightly more efficient at metallicities similar
to solar ($Z \simeq 0.02$) than it is at lower metallicities $Z \lesssim 0.005$.
There is no one single reason for this behaviour, and it is most likely a combination of small effects 
such as different radii and different wind mass loss rates of stars with different metallicities. 
We thus consider this result to be much less robust than the increasing formation efficiency of BH-BH mergers towards lower metallicity.

The updated initial binary distributions 
of \citetalias{md16} (green lines, I1) result in smaller formation efficiencies of CBM across 
all the metallicities with respect to the previously adopted distributions of \citet[][blue lines, M10]{Sana2012}.
The only exception is the case of BH-NS mergers at $Z \approx \zsun$, where the formation efficiency
$X_{\rm BHNS} \; (\msun^{-1})$ is at its lowest and there are few such systems in our simulations (fewer than 20). 
On the other hand, the inclusion of a top-heavy IMF for $Z < 0.004 = 0.2\zsun$ (red lines, I2) significantly increases 
the formation efficiency of BH-BH and BH-NS mergers
by up to nearly an order of magnitude for the lowest $Z = 0.0001 = 0.005\zsun$ in the case of BH-BH.  The
NS-NS mergers originate from binaries with less massive primaries, which is why there is little change between 
models I1 and I2 in this case.

\section{Merger rates and LIGO/Virgo predictions}

\label{sec:res_merger_rates}
\subsection{Source-frame merger rate density}

\begin{figure}
        \includegraphics[width=\columnwidth]{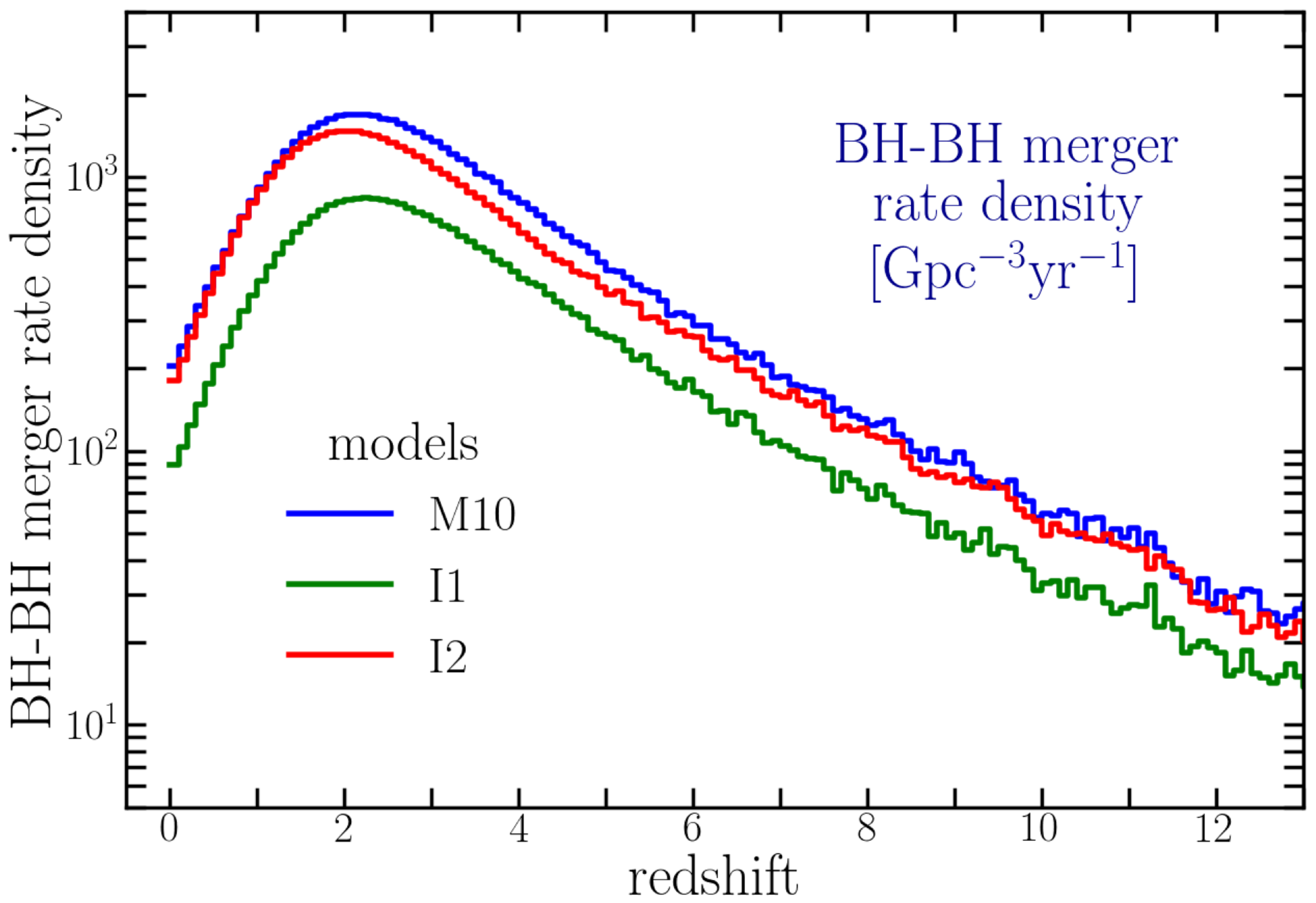} \\ 
        \includegraphics[width=\columnwidth]{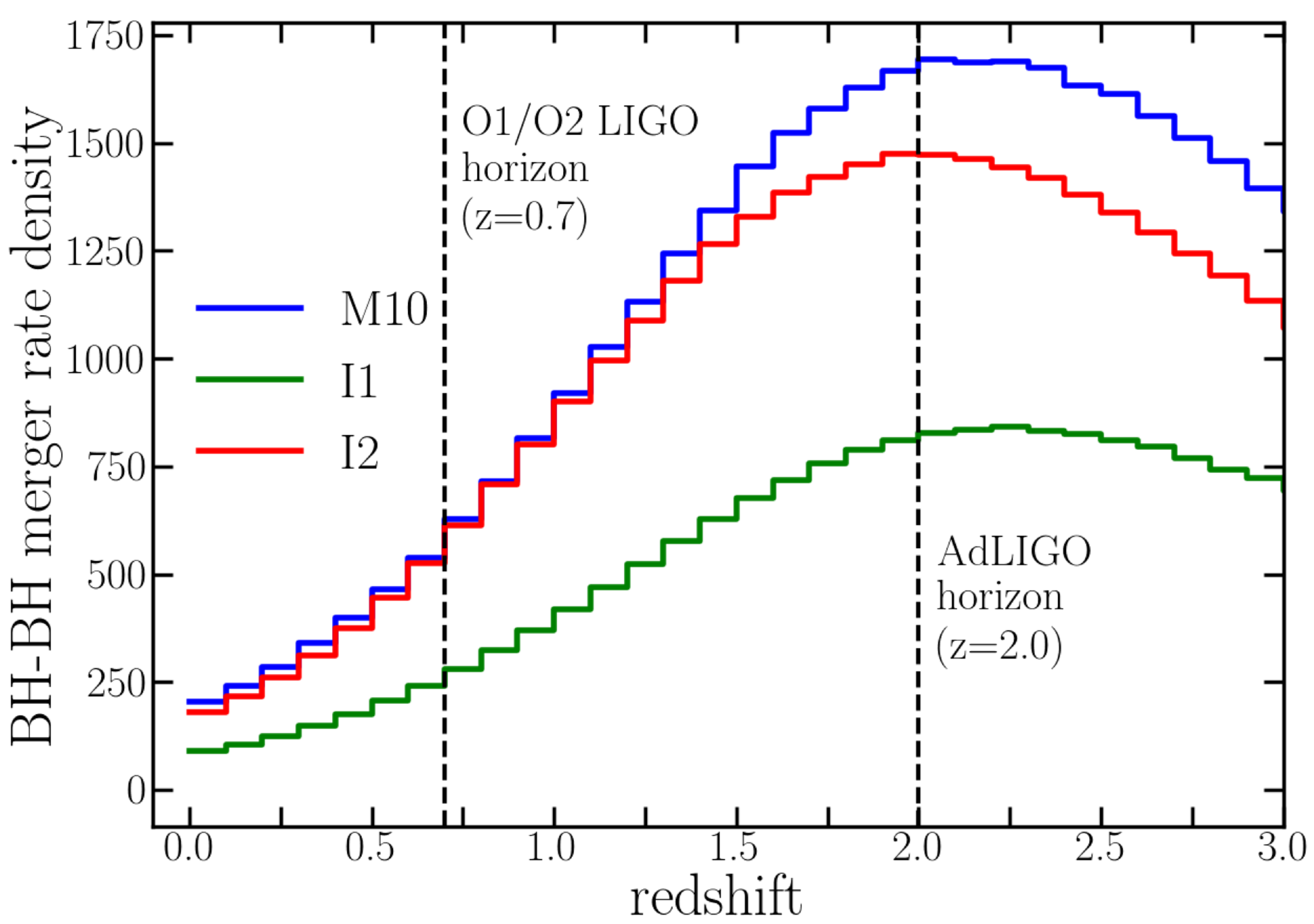}
    \caption{Source frame BH-BH merger rate density $Gpc^{-3} \, yr^{-1}$ as a function of redshift 
      for our three models: M10 (binary distributions of \citet{Sana2012}), I1 (binary distributions of \citetalias{md16})
,      and I2 (same as I1 but with a metallicity-dependent IMF). \textit{Upper panel:} the peak around $z \approx 2$ in all 
      the models corresponds to the maximum in the SFR (see text). \textit{Lower panel:} 
      the range of redshifts accessible with the current and future sensitivity of LIGO detectors. 
      We indicate the horizon redshifts for observation of an optimally inclined BH-BH merger with a total mass of $\sim 80 \msun$ (roughly the most massive systems in our simulations)
      for the sensitivities of O1 and O2 observing runs , and for the Advanced LIGO (AdLIGO). 
      The detection distances for NS-NS mergers \citep{Chen2017} were about $d_{\rm NSNS} = 70$ Mpc for O1 and O2 sensitivity \citep{LVK_prospects_2018}.
      The local BH-BH merger rate density (within $z < 0.02$) in our models 
      are about 203 (M10), 89 (I1), and 181 $\gpy$ (I2).
      The current limits imposed by the existing LIGO detections are 12$-$219$\gpy$ \citep{LIGO_GW170104}.}
\label{fig.mrd}
\end{figure}

In Fig.~\ref{fig.mrd} we show the source frame BH-BH merger rate density ($Gpc^{-3} \, yr^{-1}$)
as a function of redshift for our three models.  We note that the results for M10 are also corrected with respect to \citet{Belczynski2016_PIP} for a typo in the calculation 
of simulation normalization, yielding a small net decrease of merger rates by a factor of 0.926 (Sect.~\ref{sec.differences_wrt_B16}).

The behaviour of the merger rate density relation with redshift (upper plot) closely resembles the star formation history \citep{Madau2014}and peaks around $z \approx 2$. 
This results from the fact that the distribution of merger delay times of close BH-BH binaries follows a power-law $\propto t_{\rm del}^{-1}$ 
\citep{Dominik2012,B16Nat}, and most of the systems merge relatively shortly after their formation (a few hundred Myr). We note that the maximum 
of BH-BH merger rate density in model I2 is very slightly 
($\Delta z \approx 0.2$) shifted towards smaller redshifts. This is because in this variation, at each redshift, we assume 
a different IMF based on the mean metallicity at that redshift, and consequently apply 
the adequate IMF-SFR correction factor (see Appendix~\ref{sec:app_conv_coeff} for details). As a result, the
location of the SFR maximum is slightly different.

The lower panel shows the 
range of redshifts up to $z = 3.0$. 
We also indicate the horizon redshifts for observation of an optimally inclined BH-BH merger with a total mass of $\sim 80 \msun$ (roughly the most massive systems in our simulations)
for the sensitivities of O1 and O2 observing runs ($z_{\rm horizon} \approx 0.7$), and for the design sensitivity of Advanced LIGO
\citep[$z_{\rm horizon} \approx 2.0$;][]{LVK_prospects_2018}. For comparison, the detection distances for NS-NS mergers \citep[$d_{\rm NSNS}$; i.e. the radius of a sphere 
with an equal volume to the peanut-shaped LIGO response function; see ][]{Chen2017} were about $d_{\rm NSNS} = 70$ Mpc for O1 and O2 sensitivity.
The local BH-BH merger rate density (within $z < 0.02$) in our models are about 203 (M10), 89 (I1), and 181 $\gpy$ (I2), 
which are all within the most up-to-date range determined by the existing LIGO detections \citep[12$-$219$\gpy$][]{LIGO_GW170104}. 

\subsection{LIGO/Virgo detection rates}
\label{sec:res_ligo_det_rates}

In Table~\ref{tab.det_rates}, we summarize our predictions for the local (i.e. within $z < 0.02$) 
merger rate densities, detection rates ($R_{\rm det} \, yr^{-1}$) with the sensitivity of O1/O2 LIGO/Virgo observing runs,
and corresponding numbers of detections assuming 120 days of coincident data acquisition.
The inclusion of the updated initial MS binary distributions of \citetalias{md16} (model I1) has led to 
a decrease in both the local BH-BH merger rate density and the detection rate during O2 by a factor of $\sim$2.3
(with respect to model M10).
The assumption of a metallicity-dependent IMF (model I2, see Eq.~\ref{eq:IMF_Z}), on the other hand, causes an increase 
in the local BH-BH merger rate density by a factor of $\sim2$, and an increase in the O2 detection rate by a larger 
factor of $\sim2.5$ (with respect to model I1). The difference between these two factors arises from the fact that 
the top-heavy IMF for low metallicities assumed in I2 favours more massive BHs, which are easier to detect.
We discuss the comparison of our results with LIGO/Virgo detection rates in Sect.~\ref{sec:rates_comparison}.

\subsection{Metallicity distribution of BH-BH progenitors}

\begin{figure*}
\centering
        \includegraphics[width=0.8\textwidth]{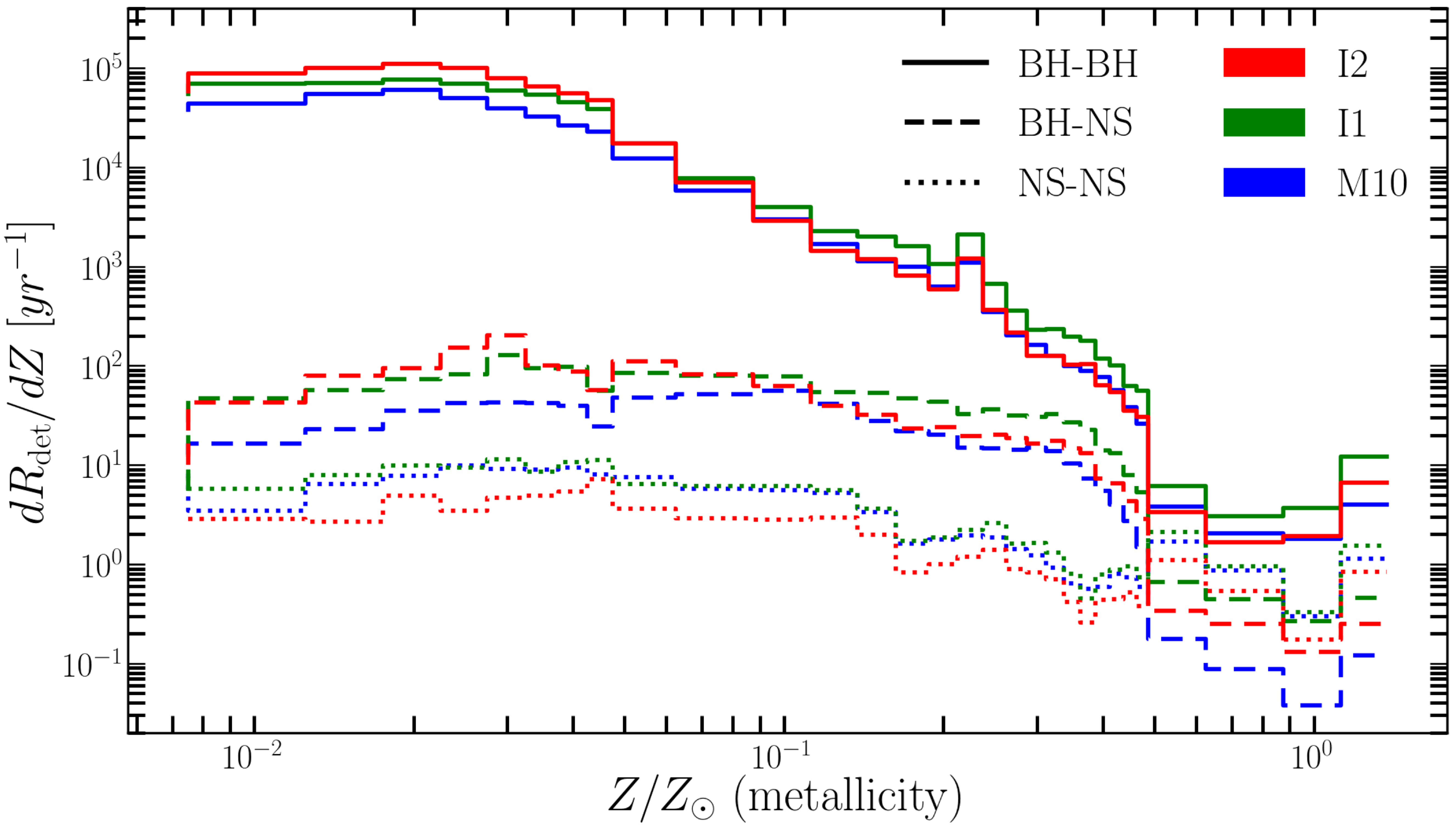}
    \caption{Detector-frame distribution of metallicities of BH-BH, BH-NS, and NS-NS merger progenitors.
    More than 90\% of BH-BH merger detections in all our models
    are systems formed with metallicites lower than 10\% of the solar metallicity, i.e. $Z<0.1\zsun$. 
    Such a strong dependence of compact binary merger formation with metallicity (see also Fig.~\ref{fig:Xcmb}),
    especially in the case of BH-BH systems, indicates the significance of SFRs at different 
    metallicites across the Universe for the compact binary merger rates (see discussion 
    in Sect.~\ref{sec:disc_cosmos}).}
\label{fig.det_frame_ZZ}
\end{figure*}

In Fig.~\ref{fig.det_frame_ZZ}, we show the metallicity distributions
of different types of CBM weighted by their detection rates with O1/O2 sensitivity.
In the case of BH-BH mergers, the huge preference for very low metallicities ($Z \lesssim 0.001 = 0.05\zsun$) 
with respect to metallicities close to solar ($Z \gtrsim 0.01 = 0.5\zsun$) 
is already present when looking at the formation efficiencies across $Z$ alone (see Fig.~\ref{fig:Xcmb}). 
At that stage, which does not account for any cosmological calculations yet (i.e. SFR, cosmic metallicity distribution, 
BH-BH merger delay times), 
there are about 2 orders of magnitude more BH-BH mergers formed with low $Z \lesssim 0.001$ compared to $Z \sim \zsun$
(solely due to binary evolution effects).
Notably, this difference grows much higher in the detector frame, with the values of d$R_{\rm det}/$d$Z$ being 
about 4 order of magnitude larger for $Z \lesssim 0.001$ than for $Z \sim \zsun$. 
This is a combination of two effects. First, at low metallicities BHs are born more massive and their mergers are easier 
to detect with the LIGO/Virgo sensitivity. Second, in our approach to population synthesis on cosmological scales
(Sect.~\ref{sec:cosmo_rates_method})
the metallicity distribution of stars forming at redshifts around the maximum of the cosmic SFR at $z \approx 2$, from which most of the BH-BH merger progenitors originates (see Fig.~2 of \citetalias{B16Nat}),
is centred at $Z < 0.1\zsun$ (see Fig.~6 of Extended Data of \citetalias{B16Nat}).

\begin{table}
\caption{Merger rate densities and detection rates for the LIGO/Virgo O2 run sensitivity for 
three models: M10 (initial binary distributions from \citet{Sana2012}), I1 (updated distributions 
from \citetalias{md16}), and I2 (I1 plus top-heavy IMF for low metallicities from \citet{Marks2012b}).}                
\centering          
\begin{tabular}{c| c c c}     
\hline\hline       
Model & Rate density\tablefootmark{a} & O2 rate\tablefootmark{b} & O2\tablefootmark{c}\\ 
merger type   & [Gpc$^{-3}$ yr$^{-1}$] & [yr$^{-1}$] & [120 days] \\ 
\hline          
   M10 &&&\\
   NS-NS &  68.2 &   0.09 &  0.03 \\  
   BH-NS &  26.7 &   0.42 &  0.22 \\
   BH-BH &   203 &    111.6 &  36.7 \\
\hline                  
   I1 &&&\\
   NS-NS &  25.9 &   0.03 &  0.01 \\  
   BH-NS &  13.3 &   0.25 &  0.08 \\
   BH-BH &   88.8 &    48.6 &  16 \\
\hline                  
   I2 &&&\\
   NS-NS &  24.4 &   0.03 &  0.01 \\  
   BH-NS &  16.5 &   0.38 &  0.13 \\
   BH-BH &   181 &    122.7 &   40.3 \\
\hline \hline        

\end{tabular}
\tablefoot{\\
\tablefootmark{a}{Local merger rate density within redshift $z<0.1$.}\\
\tablefootmark{b}{Detection rate for LIGO O2 observational run.}\\
\tablefootmark{c}{Number of LIGO detections for effective observation time in O2.\\}
}
\label{tab.det_rates}   
\end{table}

\section{Discussion}

\label{sec:discussion}

\subsection{Understanding the impact of the new initial distributions}
\label{sec:disc_understanding}
Within the classical binary evolution scenario,
our predictions for the merger rates of double compact objects based on the 
updated initial MS binary parameters from \citetalias{md16}
are $\sim2-2.5$ times smaller than the results previously obtained by \citetalias{dMB15} and \citetalias{B16Nat}
from the distributions of \citet{Sana2012}. This can be explained 
by looking at the differences between the two distributions
within the areas of the parameter space that are most relevant 
for the formation of double compact object mergers. Let us first 
take a look at BH-BH and NS-NS progenitors.

(i) In agreement with \citetalias{dMB15}, we find that the majority 
of BH-BH and NS-NS mergers originate from binaries with
initial periods within $\logpday$ = 2$-$4. The fraction 
of primaries residing in such systems is about 
F$_{\rm logP=2-4;bin}$ = 28\% in the case of \citet{Sana2012} distributions.
Meanwhile, this value is higher by about $\sim1.35$
in the case of statistics obtained by \citetalias{md16} (Fig.~\ref{fig:pdist}): i.e.
F$_{\rm logP=2-4;bin}$ = 37\% for $M_1$ = $8 \msun$ primaries,
and F$_{\rm logP=2-4;bin}$ = 40\% for $M_1$ = $30 \msun$ primaries
(regimes for NS-NS and BH-BH progenitors, respectively).

(ii) The eccentricity distribution based on the \citet{Sana2012}
sample is skewed towards lower values with $<$\,$e$\,$>$ = 0.3
compared to the statistics reported by \citetalias{md16} 
with $<$\,$e$\,$>$ = 0.6 in the case of massive binaries (Fig.~\ref{fig:edist}).
The progenitors of CBM derive from binaries 
with periastron separations log $d_{\rm per}$ ($\rsun$) = 2$-$3.5, 
depending slightly on metallicity and the exact formation channel.
By increasing the average initial eccentricity from $<$\,$e$\,$>$ = 0.3 to 0.6, then the average
orbital separation of the progenitors increases by a factor of (1$-$0.3)/(1$-$0.6) = 1.75.
This corresponds to a shift in logarithmic orbital
period of $\Delta$log$P$ $\approx$ 0.2. Considering the progenitors derive
from a broad parameter space of orbital periods $\logpday$ = 2$-$4, 
a slight shift of $\Delta$log$P$ $\approx$ 0.2 due to the updated eccentricity
distribution does not affect the predicted rates of compact object mergers.  

(iii) Similar to \citetalias{dMB15}, we find that nearly all BH-BH
and NS-NS mergers derive from initial MS binaries 
with mass ratios $q$ = 0.5$-$1.0. According to the old
mass-ratio distribution from \citet{Sana2012}, $\approx$55\% of early-type binaries
have mass ratios across this interval.
According to the updated period-dependent mass-ratio distribution of \citetalias{md16},
however, only (17-28)\% of massive binaries with $\logpday$ = 2$-$4
have mass ratios $q$ = 0.5$-$1.0, which results in about $\sim$2.0$-$3.2 times
fewer potential BH-BH and NS-NS merger progenitors.

Altogether, the new mass ratio distribution that is shifted towards smaller values with respect to the old distribution (iii) is the 
most significant update in the context of BH-BH and NS-NS mergers, and the differences between period distributions are of secondary importance (i). Based 
on the estimates presented above, we could expect 
that there should be roughly between 1.5 and 2.4 fewer BH-BH and NS-NS mergers 
in the updated simulations. This explains the results of our computations (see tables~\ref{tab:ZAMS_mass_fractions}
and~\ref{tab.det_rates}).

In the case of BH-NS mergers, the binary parameters of their 
progenitors depend strongly on metallicity (Fig.~\ref{fig:BHNS_properties}). 
Around solar metallicity, there are few to no BH-NS systems formed 
and our sample is not statistically significant.
At subsolar metallicities of $\sim0.1\zsun$,
the main formation channel involves two CE phases, and
this channel dominates when summed across all metallicities. 
It requires the initial mass ratios between 0.3 and 0.6 (peaking around 0.4) and, more 
importantly, pericentre separations 
of log $d_{\rm per}$ ($\rsun$) = 3.2$-$4, preferably over 3.5. The updated initial distributions predict fewer massive binaries
this wide. Additionally, in most cases the NS is formed from a star
with mass at the higher end of the mass range of NS progenitors ($15 \lesssim M_2/\msun \lesssim 22$), 
which helps to produce a high enough mass ratio at a BH-MS binary stage, so that a dynamically 
unstable mass transfer occurs and a CE phase is triggered. The shift towards smaller mass ratios
in the updated distributions causes a corresponding shift towards slightly more massive primaries.
This additionally decreases the BH-NS formation efficiency due to a declining slope of the IMF. \\

At very low $Z$ $\sim$ $0.01\zsun$, the dominant formation 
channel involves binaries with components of initially similar masses 
that evolve into double-giant systems and experience a stable mass transfer 
instead of the first CE (Sect.~\ref{sec:bhns_prog}).
The requirement of $q$ > 0.9 means that there are about three times fewer potential 
progenitors for this channel in the updated distributions.
With all the metallicities combined, all the factors described above result 
in the merger rate of BH-NS systems being about two times smaller 
in the new simulations (Table~\ref{tab.det_rates}).

\subsection{Small impact of a top-heavy IMF}
\label{sec:small_imp}
The uncertainty in the IMF is often considered to have a substantial impact on the compact 
binary merger rates, by a factor of a few \citepalias{dMB15} up to an order of magnitude 
\citep[see Fig.~20 of ][]{Kruckow2018}.
One could thus expect that the top-heavy IMF for low metallicities 
($Z<0.2\zsun$; see Eqn.~\ref{eq:IMF_Z}) we implemented in model I2
would result in a significant increase of BH-BH merger rates. 
According to Fig.~\ref{fig.det_frame_ZZ}, the vast majority of BH-BH mergers 
detectable with LIGO/Virgo were formed in $Z<0.1\zsun$ metallicity environments. In model I2, 
the IMF power-law exponent for massive stars is $\alpha_3 \approx 2.07$ 
for $Z=0.1\zsun$ decreasing down to as low as $\alpha_3 \approx 1.31$ for $Z=0.01\zsun$.
For a fixed amount of stellar mass formed (i.e. fixed SFR) and with respect to the standard value $\alpha_3 = 2.3$ as in 
models M10 and I1, this would correspond to an increase in the number of primaries within 
$\rm M_1$ = 45$-$55 $\msun$ (highest bin in the primary mass distribution 
of BH-BH progenitors, Fig.~\ref{fig.ZAMS_properties}) by a factor of about $\sim2.2$ 
at $Z=0.1\zsun$ and $\sim5$ at $Z=0.01\zsun$.
Meanwhile, the merger rate density of BH-BH in model I2 (with a top-heavy IMF) is higher than in model I1 by only a factor
of $\sim 2$. 

This showcases the significance of keeping the consistency between an adopted 
cosmic SFR and an assumed IMF. Because the cosmic SFR is measured based on UV observations 
\citep{Madau2014}, we normalize our simulations in such a way that the total amount 
of UV light at $\rm 1500 \AA$ stays roughly constant, irrespective of the assumed IMF
(see Appendix~\ref{sec:app_conv_coeff}). As most of the UV light is produced by massive stars, 
a top-heavy IMF is characterized by a higher UV luminosity per unit of stellar mass formed $\mathcal{U} \; (\msun^{-1})$ 
and must result in a rescaling of the SFR to a smaller value.
A smaller SFR means fewer stars formed, which counteracts 
the fact that with a top-heavy IMF a higher fraction of these stars are massive.
A similar but inverted argument could be made for a top-light IMF (i.e. $\alpha_3>2.3$). 
Thus, the total number of massive stars formed stays roughly constant irrespective of the IMF variations, 
as it is constrained by the UV observations.

\begin{figure}
\centering
    \includegraphics[width=\columnwidth]{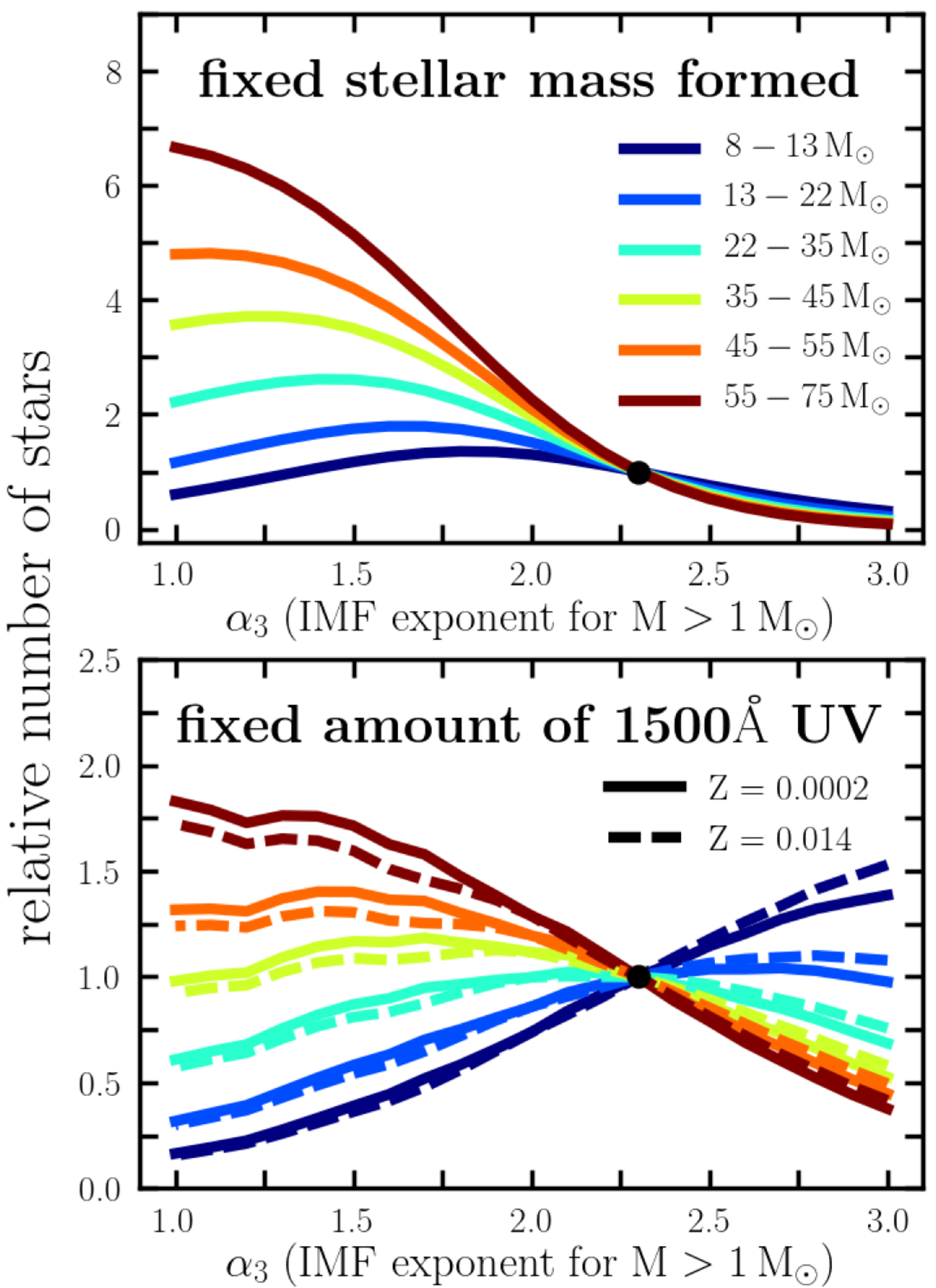}
    \caption{Relative numbers of massive stars $\mathcal{R}$ formed within multiple initial mass ranges as 
    a function of power-law exponent $\alpha_3$ for the high-mass IMF component ($\rm M > 1 \, M_{\odot}$).
    See eq.~\ref{eq:relative_numbers} for the exact definition of $\mathcal{R}$.
    \textit{Upper panel}: the same SFR, irrespective of the IMF changes. The number 
    of stars is thus calculated with respect to a fixed stellar mass formed. 
    \textit{Lower panel}: SFR corrected for the IMF changes (see Eqn.~\ref{eq:SFR_conversion}). 
    The number of stars is thus calculated with respect to a fixed amount of UV radiation 
    at $1500\AA$ wavelength that is constrained by the observations \citep{Madau2014}. }
\label{fig.relative_numbers}
\end{figure}

We show this quantitatively in Fig.~\ref{fig.relative_numbers}, where, as a function of 
$\alpha_3$, we plot the relative number $\mathcal{R}(\alpha_3)$ of stars formed in a given mass bin $M_{\rm low}$-$M_{\rm high}$ (e.g. between 8 and 13 $\msun$, and so on) defined as
\begin{equation}
\label{eq:relative_numbers}
 \mathcal{R}(\alpha_3) = \frac{M_{\rm total;1}(\alpha_3) \times \int_{M_{\rm low}}^{M_{\rm high}}\xi(\alpha_3,M) dM}{M_{\rm total;2}(\alpha_3 = 2.3) \times \int_{M_{\rm low}}^{M_{\rm high}} \xi(\alpha_3 = 2.3,M) dM}
,\end{equation}
where $\xi(\alpha_3,M)$ is a Kroupa-like IMF in the form given by
Eq.~\ref{eq:IMF} for which we vary the high-mass slope $\alpha_3$, 
and $M_{\rm total;1}$ and $M_{\rm total;2}$ are the total amounts of stellar mass formed in models with different IMFs.
The IMFs are normalized to unity, i.e. $\int_{0.08}^{150} \xi(\alpha_3,M) dM = 1.0$ for any slope $\alpha_3$.
The upper panel represents simulations normalized to a fixed amount of stellar mass formed,
irrespective of the IMF (the usually made assumption), so $M_{\rm total;1} = M_{\rm total;2}$.
The lower panel, in turn, assumes a fixed UV luminosity of a star-forming population, 
so that $M_{\rm total;1} \times \mathcal{U}(\alpha_3) = M_{\rm total;2} \times \mathcal{U}(\alpha_3=2.3)$.
The amount of UV light per unit of star-forming mass $\mathcal{U} \; (\msun^{-1})$ is inversely proportional to $\mathcal{K}$
coefficients in Eq.~\ref{eq:app} and Table~\ref{tab:calib_coeff}.

We note that although the relative number of primaries does not change by more than a factor of $\sim 2$ in 
any of the mass ranges and IMFs shown, the ratio of progenitors of massive BH-BH to NS-NS mergers 
changes more significantly. \footnote{The simplified way we correct the SFR explains why the BH-BH merger rate increases by a factor 2 between models I1 and I2 
while none of the curves in lower panel of Fig.~\ref{fig.relative_numbers} actually reaches 
the value of 2. In practice, at each redshift bin 
we assume that the amount of UV light per unit of star-forming mass is such as for the median metallicity 
(and IMF corresponding to it) at that redshift. However, given the uncertainty of the contributions from different
metallicities to the cosmic SFR, this simplification has no meaningful impact on the results.}

\subsection{Population synthesis on cosmological scales}
\label{sec:disc_cosmos}
The discovery of GW150914 revealed the existence of stellar BHs with masses of 30\,$-$\,40$\msun$ \citep{LIGO2016_GW150914}.
Black holes this massive are only expected to form from low-metallicity stars $Z/\zsun \lesssim 0.3$ \citep{Belczynski2010a}, and this conclusion 
is strengthened by a recent update to Wolf-Rayet star winds \citep{Vink2017}.
Additionally, Fig.~\ref{fig:Xcmb} shows that the formation efficiency of BH-BH mergers is very strongly dependent on metallicity. 
For that reason, predictions for the rates of such events are sensitive to the applied distribution of metallicity of star formation throughout the Universe 
and they cannot be made by simply extrapolating results for a Milky Way-like galaxy, as still done by some authors. 

We have adopted the mean metallicity-redshift relation from the chemical evolution model of \citet[][see Eqn.~\ref{eq:Zmean}]{Madau2014}
with the level increased by 0.5 dex and with a Gaussian spread of $\sigma$ = 0.5 dex, and used this relation
to weight the contributions from different metallicities to the SFR at different redshifts.
This cosmological information, combined with the LIGO/Virgo sensitivity, allowed us to transform our simulated CBM into detector-frame predictions. 

However, the mean metallicity of the Universe (defined as the total mass of heavy elements ever produced per mass of baryons in the Universe)
in general does not equal the mean metallicity of star formation at a given redshift.
This simplification is a caveat of our model to keep in mind. The mean metallicity 
at which star formation takes place is likely higher,
since the highest SFRs are revealed by massive galaxies \citep{Lara-Lopez13}, which are also relatively
metal rich \citep[e.g.][]{Tremonti04}.
Low-metallicity star formation in the local Universe, on the other hand, occurs mostly in low-mass 
dwarf galaxies \citep{Andrews2013}, which have relatively little star formation \citep{Boogaard2018}.
For that reason, we are likely somewhat overestimating the current SFR of metal-poor massive stars. 
For example, according to our model, about 18\% of massive stars that recently formed in the local Universe (z = 0) 
have $Z < 0.1 \zsun$. Observations of nearby dwarf and spiral galaxies suggest only a few percent, perhaps at most 10\%, 
of massive stars form locally at such low metallicity (see App.~\ref{sec:sfr_calc} for a detailed calculation).
While this signifies the existence of a problem, the issue is more complicated and requires a coherent and
observationally based model of cosmic star formation at different
metallicities and across different redshifts (Chruslinska et al. in prep.).

A different approach to population synthesis on a cosmological scale is to distribute simulated systems over individual galaxies. 
This could be obtained in a semi-analytical manner by applying a galaxy mass function \citep{Elbert2018} or fits to galaxy trees from
cosmological simulations \citep{Lamberts2016}, and combining these with observationally inferred galaxy scaling relations for SFRs and metallicity distributions \citep[eg.][]{Behroozi2013}. Alternatively, results from cosmological simulations 
could be applied directly to obtain similar information about star formation in individual galaxies \citep{Schneider2017, Mapelli2017}. 
Interestingly, \citet{Mapelli2017} predict that the distribution of metallicity of BH-BH merger progenitors
peaks at around $\sim 0.1\zsun$ and declines towards lower $Z$, which conflicts with our Fig.~\ref{fig.det_frame_ZZ}.
The most likely explanation of 
this discrepancy is that we are overestimating the contribution from very low metallicities at close redshifts of star formation.
The majority of the low-metallicity content of the recent Universe is locked in dwarf galaxies with little star formation, for which
our simplification that the mean metallicity of the Universe $\approx$ mean metallicity of star formation is likely less accurate.
This interpretation also agrees with the fact that \citet{Mapelli2017} 
did not find the clearly bimodal distribution of massive BH-BH merger birth redshifts as shown in the case
of our simulations in Fig.~2 of \citetalias{B16Nat}, 
but rather only the single maximum around the peak of cosmic star formation at $z \approx 2$. 
However, even if we were overpredicting the star formation in low metallicities $Z < 0.1\zsun$ at redshifts $z < 2$ to the point at which we would have 
to discard all our mergers formed in these conditions, this would still not affect our predictions for the merger rates by more than a factor of 2.

Although making predictions for CBM based on cosmological simulations as carried out by
\citet{Schneider2017} or \citet[][or, in a simplified way, also by \citealt{Belczynski_NSNS_origin}]{Mapelli2017}
may be burdened with additional biases,
this ultimately seems to be the superior approach that hopefully will allow us to confront models with
various galactic properties in the case of mergers with identified hosts \citep[such as GW170817,][]{NSNS_discovery_paper}. 

Finally, we wish to highlight that the small impact of IMF variations on merger rates (\ref{sec:small_imp})
is a general result that should also appear in the predictions based on cosmological simulations.
Even though such simulations often provide SFR values as their output, changing the assumptions
on the massive-star IMF without renormalizing the SFR make their results inconsistent with star formation tracers such as UV observations.

\subsection{Comparison with LIGO/Virgo detection rates and other studies}
\label{sec:rates_comparison}
We show the merger ($\rm Gpc^{-3} yr^{-1}$) and detection rates ($\rm yr^{-1}$) 
from our simulations in Table~\ref{tab:calib_coeff}.
While our models perform relatively well in retrieving the BH-BH merger rate as constrained by
the LIGO/Virgo observations $R_{\rm BHBH} = $12$-$213$\gpy$ \citep{LIGO_GW170104}, 
they underpredict the rate of NS-NS mergers that was inferred after the recent detection of
GW170817: $R_{\rm NSNS} = 1540 ^{+3200}_{-1220}\gpy$ \citep{NSNS_discovery_paper} \footnote{We note that because only one NS-NS merger has been detected so far, 
the lower limit on the inferred merger rate is rather poorly defined.}.
Recently, \citet{Chruslinska2018} analysed NS-NS merger rates within a large suite
of \textsc{StarTrack} models, varying multiple binary evolution assumptions,
and found that it is, in fact, very difficult to obtain $R_{\rm NSNS}$ consistent with the LIGO/Virgo constraints. 
The only three models marginally consistent with the reported NS-NS
merger rates, which have $R_{\rm NSNS}$ values of about $380$, $450,$ and $630\gpy$, overpredict
the merger rate of double BHs, resulting in $R_{\rm BHBH}$ of about $1070$, $310,$ and $700\gpy$, respectively.
This discrepancy could perhaps be a hint that the stability criteria for mass transfer in Roche-lobe
overflowing high-mass X-ray binaries were different in the cases of BH and NS accretors, 
resulting in a stable mass transfer (i.e. avoiding CE)
for a large space of binary parameters in the case of BH accretors \citep[see also][]{Pavlovskii2017}.

Similarly, using code $\rm C_{OM}B_{IN}E$, \citet{Kruckow2018} obtained NS-NS merger rates 
of around $10 \rm \; Gpc^{-3} yr^{-1}$ in their default model; this is well below the LIGO/Virgo limits. 
Their most optimistic model yields a value of $159 \rm \; Gpc^{-3} yr^{-1}$, which could be increased 
further up to $400 \rm \; Gpc^{-3} yr^{-1}$ to be marginally consistent with the observational constraints if natal kick
velocities were reduced by half. Contrary to \citet{Chruslinska2018}, the optimistic model of \citet{Kruckow2018} 
does not overpredict the BH-BH merger rates. However, it was only calculated for a single metallicity value Z = 0.0088 = 0.44$\zsun$
above the low-metallicity regime at which the BH-BH formation is the highest.

Interestingly, using code \textsc{MOBSA} combined with the results 
of Illustris simulations \citep{Nelson2015}, \citet{Mapelli2018} were able to obtain both NS-NS and 
BH-BH simultaneously consistent with the LIGO/Virgo constraints. 
It should be noted that they did have to assume both very low natal kicks ($\sigma \simeq 15 \rm \; km \; s^{-1}$)
for all of their core-collapse supernovae and a rather high efficiency of the CE ejection ($\alpha_{\rm CE} = 5$), which is a value that usually 
calls for some additional process aiding the CE ejection.
The former assumption can to some extent be justified by the fact that NS progenitors in binary systems tend to get their envelopes stripped 
in mass transfer episodes, which leads to less-energetic supernovae and smaller mass ejections, and possibly 
natal kicks as small as $\sim 15 \rm \; km \; s^{-1}$, although likely only in the case of ultra-stripped stars
that also lose their helium envelope to a compact object accretor
\citep[][]{Tauris2015}. As not all NS progenitors in NS-NS systems are expected to be ultra-stripped (especially those that form the first NS), 
a single $\sigma \simeq 15 \rm \; km \; s^{-1}$ value for all the core collapse supernovae may be an oversimplification
\citep[for comparison, see the model of ][]{Kruckow2018}. Observationally, some Galactic close-orbit NS-NS systems show evidence of rather large natal kicks of over
$>100 \rm \; km \; s^{-1}$, while others require small kick velocities $<50 \rm \; km \; s^{-1}$ \citep[see Sect.~6.4 of ][]{Tauris2017}.

\section{Summary}

\label{sec:summary}

Observational studies have long hinted at probable correlations between 
the initial binary parameters primary mass ($M_1$), mass ratio ($q$), orbital period ($P$), and eccentricity ($e$) \citep{Abt1990,Duchene2013}, 
but hitherto the selection biases have been too large to accurately quantify the intrinsic relations. 
A recently published paper, \citet{md16}, analysed results from more than 20 massive star surveys and,
for the first time, obtained a joint probability density function $f(M_1, q, P, e)$ for the initial ZAMS binary parameters (Sect.~\ref{sec:init_distr}).
We implement this result and analyse its impact on the predictions of merger rates of double compact object 
detectable with the LIGO/Virgo interferometers that originate from the isolated evolution scenario involving a CE phase. 
Using the \textsc{StarTrack} rapid binary evolution code \citep[Sect.~\ref{sec:physical_assumptions}, ][]{Belczynski2002,Belczynski2008}, we evolve a large population of massive MS
binaries across a range of 32 metallicities from 0.005$\zsun$ to 1.5$\zsun$
with their initial parameters drawn from the interrelated distribution of \citet{md16}. We distribute 
our systems at cosmological scales according to an observationally inferred cosmic SFR and a mean metallicity-redshift relation obtained within a chemical evolution model by \citet[][see Sect.~\ref{sec:cosmo_rates_method}]{Madau2014}.
We compare our results to those obtained by \citet{dMB15} and \citet{B16Nat} who performed similar simulations 
but with the non-correlated initial binary distributions from \citet{Sana2012}.
\\
Additionally, we discuss the level of uncertainty of compact binary merger rate predictions associated 
with possible variations of the massive star IMF. We make an in-depth study of one such variation, 
according to which the IMF becomes more top heavy (i.e. flatter power-law exponent 
for massive stars) the lower the metallicity of star formation
\citep[][see Sects.~\ref{sec:var_with_Z} and \ref{sec:imf_z_dependance_introduced}]{Marks2012b}.
We make sure that along with the changes of the IMF, the amount of far-UV light at $\rm 1500 \AA$ wavelength
\citep[based on which the cosmic SFR is measured][]{Madau2014} stays the same. 
\\
We note that our merger rates are in good agreement with the LIGO/Virgo constraints in the case of BH-BH systems,
but they are strikingly too small when it comes to NS-NS mergers \citep[Sect.~\ref{sec:res_ligo_det_rates}, see also][]{Chruslinska2018}.
However, we predict that our results of the relative impact of initial distributions and IMF variations will hold
in the general case of isolated evolution channels involving a CE phase, regardless of the absolute numbers of mergers obtained.
\\
We arrive at the following conclusions:

(a) The introduction of the updated initial ZAMS binary distributions from \citetalias{md16} (model I1) decreases 
the formation efficiency and, consequently, merger and detection rates of all types of CBM 
by a factor of about 2.1\,-\,2.6 with respect to the old simulations adopting \citet{Sana2012} distributions (model M10); see Sect.~\ref{sec:res_merger_rates}.
This is a very small change in comparison to uncertainties associated with the binary evolution \citep[e.g.][]{Eldridge2016,Mapelli2017,Stevenson2017_compas,Chruslinska2018, Kruckow2018}.
\\
In the case of BH-BH and NS-NS mergers, the major factor in play is the difference in the initial mass ratio distributions.
\citet{Sana2012} obtaiedn their binary statistics based on the spectroscopic measurement of massive O-type stars. 
For that reason their sample is limited to $\logpday<$ 3.5 (spectroscopically
detectable binaries) and dominated by very short-period orbits with P $<$ 20 days, to which they 
fit a flat mass ratio distribution. However, \citetalias{md16} have shown that wider binaries 
are weighted towards considerably smaller mass ratios (Fig.~\ref{fig:qdist}). 
According to their updated distributions, across intermediate periods $\logpday$ = 2\,-\,4, 
there are $\sim2.5$ fewer systems with $q>0.5$, which is the range of orbital periods and mass ratios 
at which BH-BH and NS-NS mergers are formed (see Fig.~\ref{fig.ZAMS_properties}).
\\
Of secondary importance is the difference between the orbital period distributions
(up to 40\% more system in the orbital period range
$\logpday$ = 2\,-\,4, depending on metallicity), whereas the 
changes in the eccentricity statistics have negligible influence 
on merger rates even though an increase from $<$e$> = 0.3$ to $<$e$> = 0.6$
may seem significant at first.

(b) The introduction of a top-heavy IMF at low metallicities \citep[Sect.~\ref{sec:imf_z_dependance_introduced},][]{Marks2012b}
results in a small increase of BH-BH merger rates by only a factor of $\sim2$, 
and even less significant differences in the case of BH-NS and NS-NS mergers (Table~\ref{tab.det_rates}). 
This might seem surprising because for the values of power-law exponent $\alpha_3$
for the IMF of massive stars of $\alpha_3 \approx 1.8$ at $Z=0.05\zsun$
and even $\alpha_3 \approx 1.3$ at $Z=0.01\zsun$, we could expect an increase in 
the number of massive BH-BH progenitors (for a fixed stellar mass formed) 
by a factor of $\sim$4$-$5 with respect to the standard value $\alpha_3 = 2.3$.
Interestingly, recent spectroscopic observations of 247 massive stars in the 30 Doradus 
star formation region reveal an IMF slope of $\alpha_3 = 1.90^{+0.37}_{-0.26}$ in the range 
15\,$-$\,200$\msun$ \citep{Schneider2018}.
However, because the formation of CBM likely
occurs at cosmological scales \citep[e.g.][]{Dominik2013,Mapelli2017},
we argue that the simulations should be normalized 
to a fixed amount of UV light at wavelength $\rm 1500 \AA$ 
(used as an SFR indicator) produced by a star-forming population.
A more top-heavy IMF produces more UV light per unit of stellar mass, 
which forces a rescale of the cosmic SFR to a lower value and prevents the BH-BH merger rates 
from increasing by more than a factor of $\sim2$. This is true even for very top-heavy IMFs with $\alpha_3$ 
value as low as $1.0$ (Fig.~\ref{fig.relative_numbers}).
We conclude that the previously reported uncertainty by a factor of 6 up and down of the merger rate predictions 
due to possible variations in the IMF \citep{dMB15} is, in fact, not more significant than a factor of $\sim$2.

\begin{acknowledgements}
We thank the anonymous referee for very useful remarks.
JK gratefully acknowledges comments and pointers from JJ Eldridge, Michela Mapelli, and Gijs Nelemans. 
KB acknowledges support from the Polish National Science Center (NCN)
grants: Sonata Bis 2 DEC-2012/07/E/ST9/01360, Meastro 2015/18/A/ST9/00746,  
and Opus (2015/19/B/ST9/01099). 
MM acknowledges financial support from NASA's Einstein Postdoctoral Fellowship programme PF5-160139.
DEH was partially supported by NSF grant PHY-1708081.
They were also supported by the Kavli Institute for Cosmological
Physics at the University of Chicago through an endowment from the Kavli Foundation.
\end{acknowledgements}

\bibliographystyle{aa}
\bibliography{ligo_mergers_bib.bib}

\begin{appendix}
\section{Conversion coefficients between SFR and different IMFs}
\label{sec:app_conv_coeff}

As we describe in Sect.~\ref{sec:cosmo_rates_method} , any cosmic SFR is associated with an assumed IMF and it is inconsistent to use it in combination with 
a different IMF. We therefore introduce a conversion factor ${\cal K}_{\rm IMF}$ 
to correct the cosmic SFR of \citet[][Eqn.~\ref{eq:SFR}]{Madau2014} obtained for 
the Salpeter IMF \citep{Salpeter1955} for the purpose of combining it with a more 
realistic Kroupa-like IMF, i.e.
\begin{equation}
\label{eq:app}
 {\rm SFR}_{\rm IMF}(z) = {\cal K}_{\rm IMF} \times {\rm SFR}_{\rm Salpeter}(z)
\end{equation}
\noindent where ${\rm SFR}_{\rm Salpeter}(z)$ is given by Eqn.~\ref{eq:SFR}. We calculate ${\cal K}_{\rm IMF}$
as a relative strength of the 1500 $\AA$ line in the UV spectrum 
of a star-forming region with SFR = $1.0 \msun\, yr^{-1}$ and the Salpeter IMF with respect to the same 
SFR but with another given IMF. 

In order to compute the UV spectra 
for different IMFs, we utilize the \textsc{Starburst99} code, designed to model spectrophotometric
properties of star-forming galaxies \citep{starburst1,starburst2,starburst3}. 
We configurate \textsc{Starburst99} to use the newest evolutionary tracks 
published by the Geneva group: for solar \citep[Z = 0.014][]{Ekstrom2012} 
and subsolar metallicity \citep[Z = 0.002][]{Georgy2013}, which cover the mass range  
$M/\msun \in [0.8, 120]$. In this paper our population synthesis simulations covered 
a much larger grid of 32 different metallicities from Z = 0.0001 up to Z = 0.03. However, 
the differences in UV-to-SFR conversion factors for different metallicities are not very significant 
\citep[see Fig. 3 of\,][]{Madau2014}. Thus, we apply ${\cal K}_{\rm IMF}$ calculated for Z = 0.014 Geneva
stellar tracks in the case of the \textsc{StarTrack} models with Z $\geq$ 0.006, and apply ${\cal K}_{\rm IMF}$ calculated for Z = 0.002 
to all the other \textsc{StarTrack} models for Z < 0.006. The upper mass limit 120$\msun$ is a limitation of 
all the stellar tracks implemented in \textsc{Starburst99}. We calculate the UV spectra up to
$10^8$ yr after the beginning of a constant star formation. By that time the 1500 $\AA$ line is 
expected to have reached its asymptotic strength already \citep[see Fig. 2 of\,][]{Madau2014}.

In Table~\ref{tab:calib_coeff} we list the values of ${\cal K}_{\rm IMF}$ computed at solar and subsolar metallicity 
for multiple power-law exponents of the high-mass end of a Kroupa-like IMF. 
Even if the high-mass end of the IMF is significantly top heavy (e.g. $\alpha_3 = 1.0$), 
we find there are only $\sim$ 30\% more stars with $M_1$=45\,-\,55$\msun$ to match the same 1500$\rm \AA$ emission strength. 

We wish to mention that the \textsc{Starburst99} code does not incorporate the evolution of binaries. 
Interacting binaries, for instance, are expected to affect the emission properties of a star-forming population such as its ionizing flux or 
a UV luminosity \citep[e.g.][]{Stanway2016}. Depending on metallicity and for a Kroupa-like IMF truncated at 100$\msun$,
\citet{Eldridge2017} have calculated that binary interactions increase 
the amount of radiation $\nu L_{\nu}$ at 1500$\rm \AA$ from a star-forming population by a factor of $\sim1.2-1.25$
relative to a corresponding population of single stars (see their Table 4). We expect that the values of our conversion 
coefficients ${\cal K}$ in Table~\ref{tab:calib_coeff} could differ by a similar factor if the effect of binaries on spectral properties 
was taken into account.

\begin{table}
\caption{Conversion coefficients ${\cal K}_{\rm IMF}$ for a cosmic SFR between the Salpeter and different Kroupa-like IMFs. In the second column, we list 
metallicities corresponding to the IMF slopes $\alpha_3$ from the first column, assuming the IMF-metallicity relation given by Eq.~\ref{eq:IMF_Z} 
after \citet{Marks2012b}.}             
\label{tab:calib_coeff}      
\centering   
\begin{tabular}{c | c | c | c | c}
\hline
&& \\
IMF             & metallicity & ${\cal K}_{\rm IMF}$    & ${\cal K}_{\rm IMF}$   & relative \#stars\\
$\xi(M)$        &  Z($\alpha_3$) & Z = 0.014            &  Z =0.002             & 45\,-\,55$\msun$ \\
& inversed Eq.\ref{eq:IMF_Z} & & & Z = 0.002 \\
\hline\hline
Salpeter\tablefootmark{a} & - & 1.0 & 1.0 & 1.25\\
\hline
Kroupa\tablefootmark{b} &&&& \\
$\alpha_3 = 3.0$ & - & 2.85 & 2.27 & 0.45 \\
$\alpha_3 = 2.9$ & - & 2.26 & 1.83 & 0.51 \\
$\alpha_3 = 2.8$ & - & 1.80 & 1.48 & 0.59 \\
$\alpha_3 = 2.7$ & - & 1.42 & 1.19 & 0.68 \\
$\alpha_3 = 2.6$ & - & 1.13 & 0.95 & 0.74 \\
$\alpha_3 = 2.5$ & - & 0.90 & 0.77 & 0.83 \\
$\alpha_3 = 2.4$ & - & 0.72 & 0.62 & 0.91\\
$\alpha_3 = 2.3$ & $\geq$ 0.004 & 0.58 & 0.51 & 1.0 \\
$\alpha_3 = 2.2$ & $\sim$ 0.0029 & 0.47 & 0.42 & 1.08 \\
$\alpha_3 = 2.1$ & $\sim$ 0.0022 & 0.39 & 0.35 & 1.15\\
$\alpha_3 = 2.0$ & $\sim$ 0.0016 & 0.33 & 0.29 & 1.20 \\
$\alpha_3 = 1.9$ & $\sim$ 0.0012 & 0.28 & 0.25 & 1.25 \\
$\alpha_3 = 1.8$ & $\sim$ 0.0009 & 0.24 & 0.22 & 1.3 \\
$\alpha_3 = 1.7$ & $\sim$ 0.0006 & 0.21 & 0.20 & 1.36 \\
$\alpha_3 = 1.6$ & $\sim$ 0.0005 & 0.19 & 0.18 & 1.37 \\
$\alpha_3 = 1.5$ & $\sim$ 0.0004 & 0.18 & 0.17 & 1.41 \\
$\alpha_3 = 1.4$ & $\sim$ 0.0003 & 0.17 & 0.16 & 1.41 \\
$\alpha_3 = 1.3$ & $\sim$ 0.0002 & 0.16 & 0.15 & 1.37 \\
$\alpha_3 = 1.2$ & $\sim$ 0.00015 & 0.15 & 0.14 & 1.31 \\
$\alpha_3 = 1.1$ & 0.0001 & 0.15 & 0.14 & 1.33 \\
$\alpha_3 = 1.0$ & <0.0001 & 0.15 & 0.14 & 1.32 \\
\hline
\end{tabular}
\tablefoot{\\
\tablefootmark{a}{Single slope: $\xi(M) \propto M^{-2.35}$ for $M/\msun \in [0.1, 100]$, \citep{Salpeter1955}}\\
\tablefootmark{b}{Triple slope: $dN/dM = \xi(M) \propto  
                \begin{cases}  
                      M^{-1.3}, & \text{for }  M/M_{\odot} \in [0.08,0.5]\\
                      M^{-2.2}, & \text{for }  M/M_{\odot} \in [0.5,1.0]  \\
                      M^{-\alpha_3}, & \text{for }  M/M_{\odot} \in [1.0,150.0],
                \end{cases} $ \\ 
                where originally $\alpha_3$ = 2.7 in \citet{Kroupa1993}.} \\
}
\end{table}

\section{Local star formation rate at metallicities $Z < 0.1 \zsun$}
\label{sec:sfr_calc}

We use observational results to do a rough
estimate of the local (i.e. z $\sim$ 0)  SFR density at metallicities below $0.1 \zsun$. At close redshifts, 
star formation at $Z < 0.1 \zsun$ occurs in galaxies with mass $< 10^7 \msun$ \citep{Andrews2013}. 
The local number density of star-forming galaxies with masses between $10^6$ and $10^7 \msun$ 
can be estimated as $10^{-0.8} \; \rm Mpc^{-3}$ \citep[from extrapolating the blue fit in Fig. 15 of][]{Baldry2012}, 
while their average SFR is measured to be around $10^{-2.5} \rm \; \msun \,yr^{-1}$ \citep{Boogaard2018}.
This implies a local SFR density at $Z < 0.1 \zsun$ of $10^{-3.3} \rm \; \msun \,Mpc^{-3} \,yr^{-1}$,  
which constitutes about $\sim 3\%$ of the total SFR density at redshift z = 0 (from Eqn.~\ref{eq:SFR}). 
While the above estimate is highly uncertain, it seems unlikely that this fraction could be higher than 10\%.

\section{Completeness of simulations}

\begin{table}
\caption{Number of mergers of respective types, BH-BH, BH-NS, or NS-NS, in our samples simulated for models 
I1 (\citetalias{md16} initial binary conditions) and I2 (additionally a metallicity-dependent IMF; values given in parenthesis)
for each of the 32 metallicities we computed. Only systems expected to merge within the Hubble time are included.
We note that for metallicities $Z\geq0.004\zsun$, models I1 and I2 
are the same, as are the simulated samples, each obtained from a population of $\sim1.54 \times 10^6$ ZAMS binaries.
For low metallicities $Z<0.004\zsun$ in model I2 the number binaries at ZAMS was lower, i.e. $1 \times 10^6$.
}
\centering
 \begin{tabular}{c| c | c | c}     
\hline 
Metallicity & BH-BH ~~~~&~~~~ BH-NS ~~~~&~~~~ NS-NS\\ 
 & model I1 (I2) & model I1 (I2) & model I1 (I2)\\
\hline\hline       
0.0001  &  4364 ~ (9856) & 103 ~ (130) & 225  ~ (42) \\ 
0.0002  &  4924 ~ (7982) & 61  ~ (76)  & 92   ~ (25) \\         
0.0003  &  4724 ~ (6375) & 79  ~ (79)  & 40   ~ (13) \\         
0.0004  &  4480 ~ (5387) & 112 ~ (93)  & 78   ~ (29) \\ 
0.0005  &  4310 ~ (4762) & 156 ~ (115) & 96   ~ (20) \\  
0.0006  &  4184 ~ (3988) & 132 ~ (154) & 96   ~ (29) \\  
0.0007  &  3878 ~ (3700) & 147 ~ (150) & 110  ~ (33) \\ 
0.0008  &  3532 ~ (3087) & 165 ~ (112) & 95   ~ (43) \\
0.0009  &  3333 ~ (2685) & 135 ~ (81)  & 110  ~ (36) \\ 
0.001   &  3110 ~ (2434) & 83  ~ (71)  & 114  ~ (35) \\         
0.0015  &  2002 ~ (1339) & 199 ~ (96)  & 94   ~ (39) \\
0.002   &  1424 ~ (809)  & 194 ~ (107) & 103  ~ (37) \\ 
0.0025  &  987  ~ (474)  & 206 ~ (96)  & 116  ~ (42) \\ 
0.003   &  931  ~ (407)  & 188 ~ (73)  & 134  ~ (50) \\         
0.0035  &  1176 ~ (530)  & 214 ~ (69)  & 99   ~ (39) \\
0.004   &  1296 ~ (1296) & 212 ~ (212) & 49   ~ (49) \\
0.0045  &  962  ~ (962)  & 244 ~ (244) & 66   ~ (66) \\ 
0.005   &  622  ~ (622)  & 232 ~ (232) & 92   ~ (92) \\ 
0.0055  &  376  ~ (376)  & 275 ~ (275) & 100  ~ (100) \\        
0.006   &  288  ~ (288)  & 274 ~ (274) & 87   ~ (87) \\ 
0.0065  &  202  ~ (202)  & 285 ~ (285) & 79   ~ (79) \\ 
0.007   &  164  ~ (164)  & 280 ~ (280) & 79   ~ (79) \\ 
0.0075  &  126  ~ (126)  & 291 ~ (291) & 50   ~ (50) \\ 
0.008   &  108  ~ (108)  & 247 ~ (247) & 47   ~ (47) \\ 
0.0085  &  70   ~ (70)   & 172 ~ (172) & 57   ~ (57) \\
0.009   &  67   ~ (67)   & 148 ~ (148) & 76   ~ (76) \\
0.0095  &  45   ~ (45)   & 112 ~ (112) & 81   ~ (81) \\
0.01    &  40   ~ (40)   & 83  ~ (83)  & 90   ~ (90) \\ 
0.015   &  27   ~ (27)   & 14  ~ (14)  & 642  ~ (642) \\        
0.02    &  26   ~ (26)   & 18  ~ (18)  & 535  ~ (535) \\        
0.025   &  41   ~ (41)   & 18  ~ (18)  & 274  ~ (274) \\        
0.03    &  55   ~ (55)   & 14  ~ (14)  & 551  ~ (551) \\        
\hline\hline
\end{tabular}
\label{tab:app_no_of_DCO}
\end{table}

\end{appendix}

\end{document}